# Multichannel Contagion vs Stabilisation in Multiple Interconnected Financial Markets

Antoaneta Sergueiva[*,**,1,2]

[*]Department of Computer Science, University College London, UK

[**]Advanced Analytics Division, Bank of England, UK[3]

## Abstract

The theory of multilayer networks is in its early stages, and its development provides powerful and vital methods for understanding complex systems. Multilayer networks, in their multiplex form, have been introduced within the last three years to analysing the structure of financial systems, and existing studies have modelled and evaluated interdependencies of different type among financial institutions. The empirical studies have considered the structure as a non-interconnected multiplex rather than as an interconnected multiplex network. No mechanism of multichannel contagion has been modelled and empirically evaluated, and no multichannel stabilisation strategies for pre-emptive contagion containment have been designed. This paper formulates an interconnected multiplex structure, and a contagion mechanism among financial institutions due to bilateral exposures arising from institutions' activity within different interconnected markets that compose the overall financial market. We introduce structural measures of absolute systemic risk and resilience, and relative systemic-risk indexes. The multiple-market systemic risk and resilience allow comparing the structural (in)stability of different financial system or the same system in different periods. The relative systemic-risk indexes of institutions acting in multiple markets allow comparing the institutions according to their relative contributions to overall structural instability within the same period. Based on the contagion mechanism and systemic-risk quantification, this study designs minimum-cost stabilisation strategies that act simultaneously on different markets and their interconnections, in order to effectively contain potential contagion progressing through the overall structure. The stabilisation strategies subtly affect the emergence process of structure to adaptively build in structural resilience and achieve pre-emptive stabilisation at a minimum cost for each institution and at no cost for the system as a whole. We empirically evaluate the new approach using large regulatory databases, maintained by the Prudential Regulatory Authority (PRA) of the Bank of England, that include verified capital requirements for UK-incorporated deposit takers and investment firms and granular information on their bilateral exposures due to transactions in the fixed-income market, securities-financing market, and derivatives market. The empirical simulations of the designed multichannel stabilisation strategies confirm their capability for containing contagion. The potential for multichannel contagion through the multiplex contributes more to systemic fragility than single-channel contagion, however multichannel stabilisation also contributes more to systemic resilience than single-channel stabilisation.

---

[1] This work is supported in part by grant ISS1415\7\65 from the Royal Academy of Engineering.

[2] We thank Paul Robinson, Oliver Burrows, David Bholat, Jean-Pierre Zigrand, Mark Flood, Alexander Lipton, Yaacov Mutnikas, Jamie Coen, Dror Kennet, Jonathan Bridges, Cian O'Neill, Murray Stephen and Veselin Karadotchev, for their support, time, advice on clarifying methodologies and prioritising results, and for coordinating access to data. We would like to thank for their feedback the attendants of the following events where this work has been presented: internal seminars and the Data World Conference at the Bank of England in 2016; the 8th Conference of the Irving Fisher Committee on Central Bank Statistics in 2016, the Financial Risk and Network Theory Conference in 2016, the Data for Policy Conference on Frontiers of Data Science for Government in 2016, and a research seminar of the Systemic Risk Centre at the London School of Economics in February 2017.

[3] Any views expressed are solely those of the author, and they should not be interpreted or reported as those of the Bank of England or its policy committees.

# I: Introduction

Real and engineered systems have multiple subsystems and layers of connectivity. Networks are now established as models providing insights into the structure and function of complex systems. Single-layer networks, however, are unable to address the emerging multilayer patterns of interactions and self-organisation among entities in complex systems. That challenge has called for the development of a more general framework – multilayer networks. The theory of multilayer networks is in its early stages, and a comprehensive review of recent progress is provided in Kivelä et al. (2014) and Boccaletti et al. (2014). Among existing studies, a promising mathematical framework is based on tensors and introduced by De Domenico et al. (2013, 2015). A special case of multilayer networks are multiplexes, where each layer consists of mostly the same nodes, and edges within a layer exist only between different nodes while links between layers exist only between instances of the same node in different layers. According to the formal definition in De Domenico et al. (2013, 2015) and Kivelä et al. (2014), a fundamental aspect of modelling multiplex networks is taking into account and quantifying the interconnectivity between layers, as it is responsible for the emergence of new structural and dynamical phenomena in multiplex networks.

Multilayer networks, through the special case of multiplexes, have only been used in the last three years to study interdependencies among entities within financial systems. Multiplexes can model different type of relations (edges) existing among a set of entities (nodes) in a system and include interlayer dependence (edges). Serguieva (2012) argued that though single-layer network models had been gradually adopted in the structural analysis of financial systems, such analysis rather required more effective models as network of networks and ensemble networks. Serguieva (2013a, 2013b) outlined how an interconnected multiplex can be used to model the different type of exposures among banks, arising from their activities in different markets trading different financial instruments, and suggested using the tensorial framework. The current paper starts with this earlier idea, and now – having access to data – develops the model in detail, implements empirically, and extends the methodology towards contagion and stabilisation analysis. Serguieva (2015, 2016) address how the multilayer network can be extended further to incorporate financial market infrastructures. A multiplex model is also used in Bargigli et al. (2015) to present the Italian interbank market, where exposures are broken down in different layers by maturity and by the secured and unsecured nature of contracts. They evaluate similarity between the structures of different layers and find the differences are significant. The conclusion is that the structural differences will have implication for systemic risk. The authors do not formulate or evaluate systemic risk, and the study considers the layers separately as a non-interconnected multiplex. The interconnected multilayer structure of the interbank market is not analysed.

Next, Poledna et al. (2015) use a multiplex model to quantify the contributions to systemic risk of the Mexican banking system from four layers: deposits and loans, securities cross-holdings, derivatives, and foreign exchange. They implement Debt Rank (Battiston et al., 2012) to measure systemic risk as fraction of the economic value in a network that is potentially affected by the distress of some banks. The systemic risk of a layer is the average Debt Rank of all banks due to their connectivity in that layer, and the total risk of the system is the average Debt Rank of all banks due to the connectivity in the projection of all layers. The results show a non-linear effect, with the sum of systemic risk of all layers underestimating the total risk. The suggested comprehensive approach in the study accounts for the capital, assets and liabilities of banks, but does not consider their minimum capital requirements and risk-weighted assets. A bank is considered failed, however, when its capital depletes to the level of minimum capital requirements, not when it depletes entirely. The minimum capital requirements are based on risk weighted assets, and two banks with the same amounts of capital, assets and liabilities, will differ in their amounts of risk-weighted assets, and therefore differ in their minimum capital requirements and their available funds to cover the liabilities. Our study shows that this requires modifying to a different extent the impacts among different financial institutions, in order to simulate contagion that accounts for each institution's individual conditions for failure and corresponding spreading rates within the contagion process. This has a significant effect on potential contagion processes and their outcome. Further, Poledna et al. (2015) consider different layers but assume the combined system is the projection of all layers rather than the multiplex of interconnected layers, and therefore do not model contagion through the multiplex structure.

The current paper also builds on research done at the Bank of England by Langfield et al. (2014), where the authors argue that markets for different financial instruments are distinct in their economic rationale and function, and discuss potential advantages of analysing the interbank market as an interlinked structure of different network layers. They provide an in-depth empirical analysis of layers in the UK banking system, but do not model a multilayer network neither quantify systemic risk. In conclusion: (i) the theory of multilayer networks is in its infancy, (ii) there are very few studies addressing multilayer or multiplex networks when analysing the structure of financial systems, (iii) existing studies of interlinkages within banking systems have recognised their multilayer structure and modelled each layer as a network, (iv) contagion processes within each layer and within the projection of all layers have also been modelled, and the corresponding systemic risk has been quantified in monetary terms. However, (i) the system has not been modelled as an interconnected multiplex; (ii) multilayer contagion processes have not been formulated, (iii) the existing single-layer contagion models are not closely aligned with regulatory requirements, and (iv) no stabilisation strategies have been designed for pre-emptive, minimum-cost contagion containment. The tensorial mathematical framework has not been used in financial analysis. With this paper we address concerns (i)-(iv), formulate solutions, and provide empirical results. We work with the tensorial framework, and in Serguieva (2016) derive step-by-step tensors of ranks two, four, and six within the context of financial systemic risk. Providing detailed domain interpretation of the models allows Serguieva (2016, 2017a) to extend the range and scope of stress-testing scenarios. Here, we will directly use the derived tensor models and focus only on concerns (i)-(iv). Their solutions effectively formulate an approach for building-in structural stability within the banking system and resilience against potential crises. Though resilience is quantified as a structural rather than monetary measure, when built in it provides for sustaining a system's monetary value. Importantly, resilience is achieved through subtly and adaptively balancing the emergence process of structure, rather than through penalising institutions. Systemic instability is due to the emerged structure rather than being a fault of an institution. We do not recommend collecting a fund of penalties and waiting for institutions to get in distress before accessing it. Instead, containment of potential contagion is achieved pre-emptively by introducing a minimum change to the structure in each period, at a minimum cost for each institution and no cost for the system as whole.

This study explores large regulatory databases, including the extensive Banking Sector Monitoring (BSM) database maintained by the Prudential Regulatory Authority (PRA) of the Bank of England, and an in-house PRA tool for verifying the Capital

Adequacy of each reporting institution. It also explores the large granular database collected by PRA through its Bank Exposures Data Request to UK-incorporated deposit takers and significant investment firms, reporting on the firm's UK-consolidated basis. These data are exemplary of the 'Big Data' or granular data now available to the Bank of England (Bholat, 2013, 2015, 2016). The exposures data request, in particular, is tailored to the purposes of structural analyses of the UK financial system, and the data include bilateral exposures resulting from institutions' activity in the main three types of markets composing the overall financial market – fixed-income, securities-financing, and derivatives markets. The three layers in the multilayer structure we model correspond to these markets. The paper is organised as follows. Section II describes and visualizes the datasets. In Section III: (i) a single-layer contagion mechanism is formulated aligned with current regulatory requirements; then (ii) corresponding relative systemic-risk indexes of institutions and absolute measures of the layer's systemic risk or resilience are quantified; and finally (iii) a single-layer strategy is designed for building in structural resilience and evaluated empirically. Section IV: (i) formulates a multichannel contagion mechanism within the banking system due to exposures arising from banks' interactions in the three interconnected markets; (ii) quantifies corresponding multiplex systemic-impact indexes of institutions and structural systemic risk of the multilayer system; (iii) designs and empirically evaluates minimum-cost multichannel stabilisation strategies. Finally, Section V states the conclusions and sets directions for further research.

## II: Empirical data and visualisation

The data used in this paper are large counterparty exposures reported by systemically important UK-incorporated deposit takers and investment firms to the Bank of England's supervisory arm, the Prudential Regulation Authority. At the time of our investigation, the data spanned five quarters, a pilot in June 2014 and collections in December 2014 to September 2015. We access from the database, the firms' twenty largest exposures to banks, where banks are broadly defined as

- banks
- building societies
- broker-dealers
- and additionally, exposures to the eight largest UK banks are reported if not a top twenty counterparty

The firms report these large exposures gross, except where a legally enforceable netting agreement exists between the transacting entities. The reports are on a UK-consolidated basis. Further, we have data on counterparty exposures broken down by financial market. Each market in turn consists of a range of financial instruments and transactions. These markets and their attendant instruments and transactions are as follows:

- the fixed income market, consisting of senior, subordinated and secured debt instruments reported gross at mark-to-market (MtM) values, further segmented by residual maturity and currency
- the securities financing market, consisting of securities lending and borrowing, and repo and reverse repo transactions reported gross notional, with further breakdowns by residual maturity, currency and type of collateral

- derivative exposures reported net MtM after collateral and net MtM at default, split by various derivative contract types

**Figure 1:** *Large exposures of UK-incorporated deposit takers and significant investment firms – empirical multilayer structure by type of market*

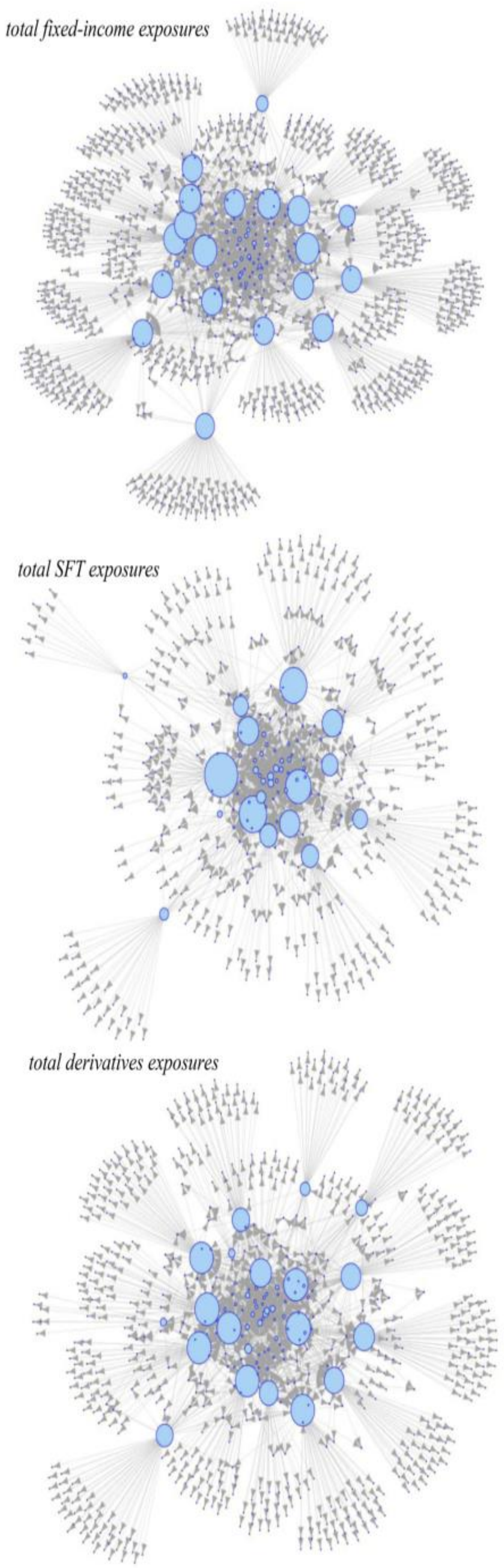

The second database used in this study is the extensive Banking Sector Monitoring (BSM) database maintained by the PRA, where we access quarterly data on UK-consolidation basis for the reporting institutions, including:

- Total Own Funds
  (Common Equity Tier 1 Capital + Additional Tier 1 Capital + Tier 2 Capital);
- Total Risk Exposure Amount (risk-weighted assets)
- Ratio of Total Own Funds to Total Risk Exposure Amount

These data are further complemented with calculations from an in-house PRA tool for verifying the Capital Adequacy of each reporting institution, including:

- Minimum Capital Requirement;
- Ratio of Available Regulatory Capital to Total Own Funds.

The empirical data on inter-institutional exposures are visualised in Figure 1, where each of the three layers corresponds to the exposure structure within a different type of market – fixed-income, securities-financing, and derivatives. The size of nodes representing institutions is proportionate to the number of exposure links they participate in. Figure 1 is based on one of the quarterly periods, between June 2014 and September 2015, however it presents key features observed in all periods – the markets differ in their emerging exposure structures. Particularly, different institutions to a different extent, and a different number of institutions, have a key role (visualised as more interconnected, larger size nodes) in different markets.

**Figure 2:** *Large exposures of UK-incorporated deposit takers and significant investment firms – empirical single network*

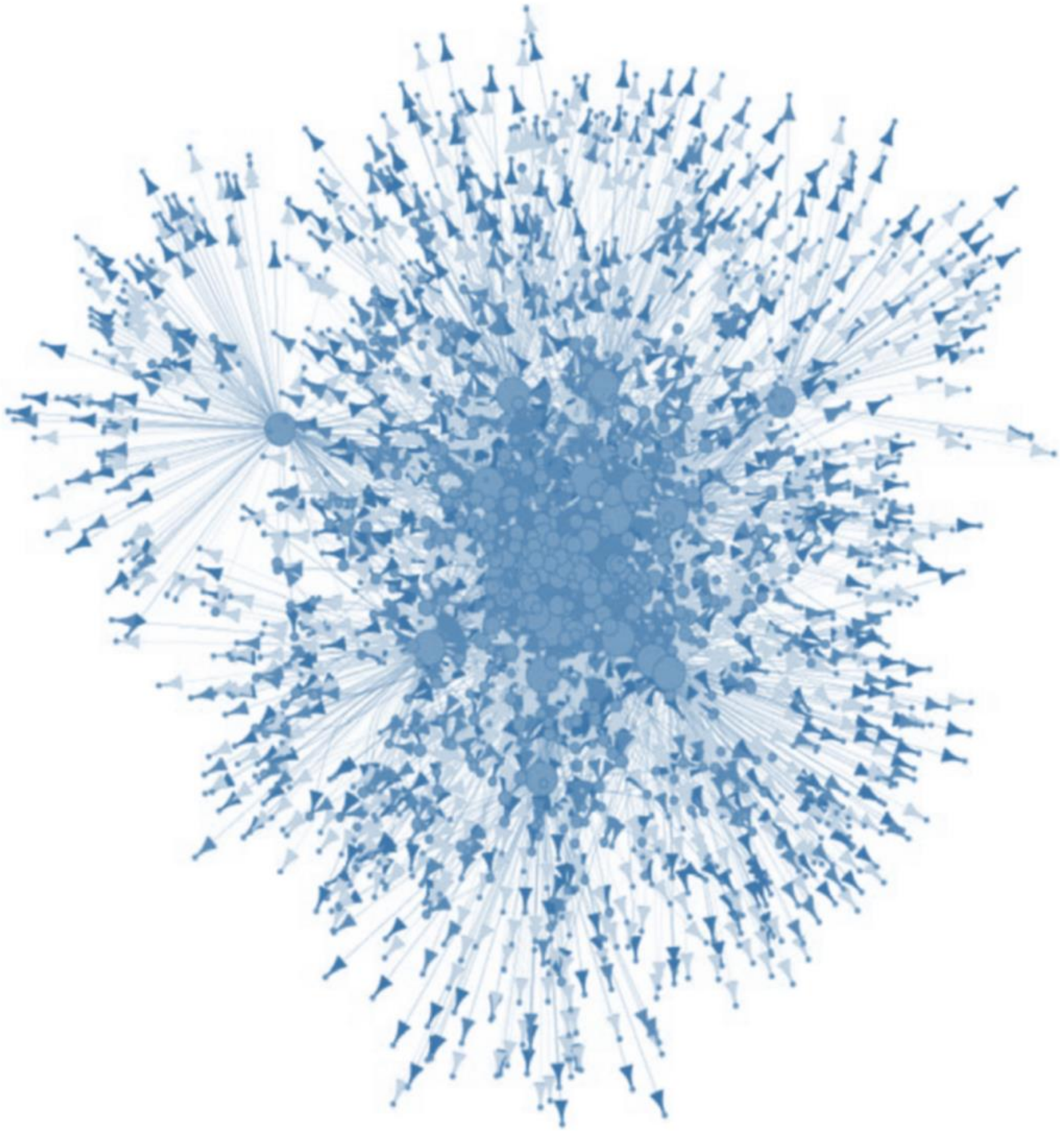

**Figure 3:** *Large exposures of UK-incorporated deposit takers and significant investment firms – empirical betweenness communities by type of market*

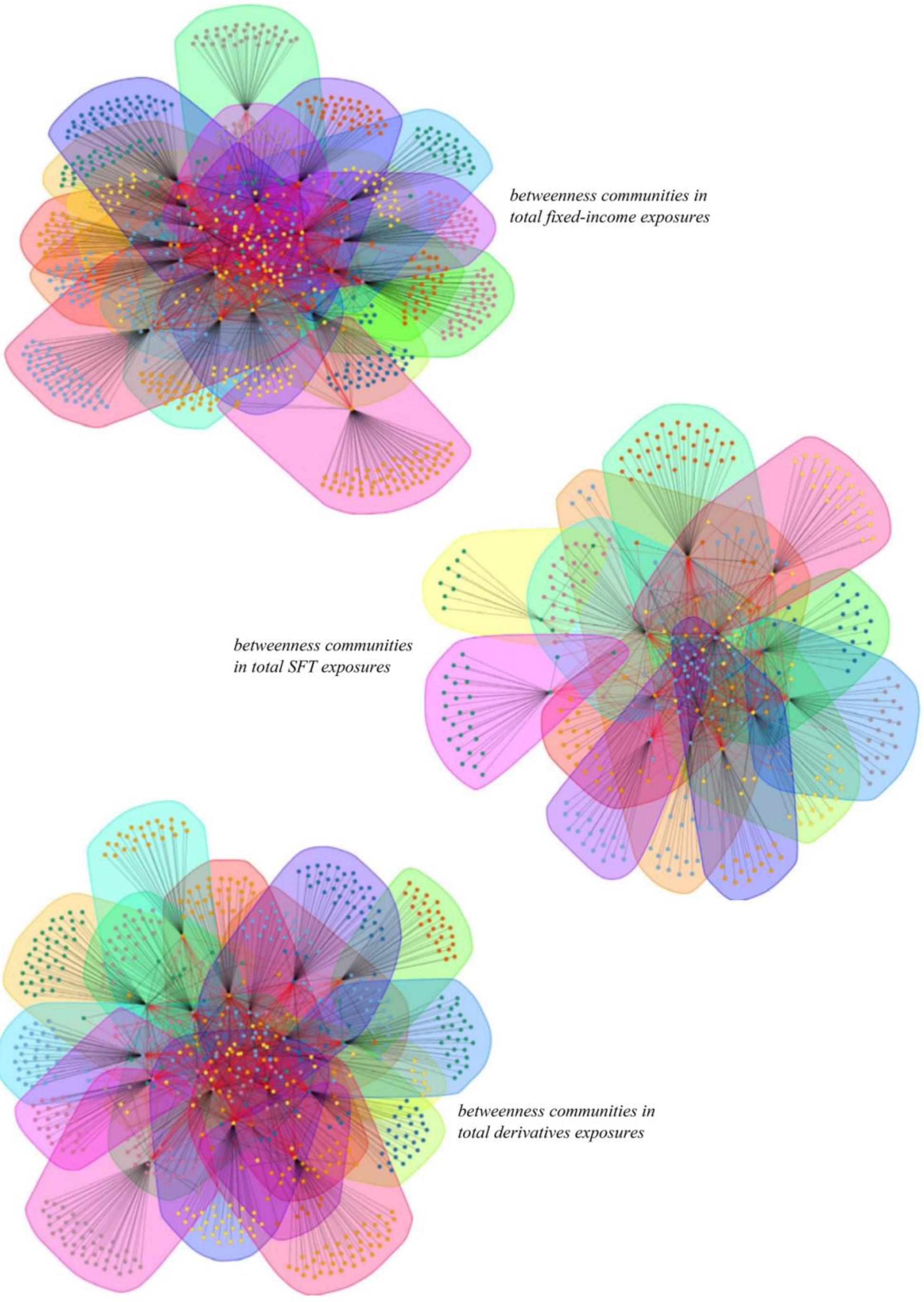

Therefore, the analysis will better inform and facilitate regulation if each market is incorporated distinctly within an overall multilayer structure, rather than all markets being amalgamated into (projected on) a single network of exposures as visualised in Figure 2. This figure presents the same quarterly period but does not observe the richer structure from Figure 1.

The argument for the structural differences between markets is further supported with the visualisation in Figure 3, where each market is clustered into communities according to edge betweenness. Betweenness of an edge (exposure link) is a measure based on the number of shortest paths (smallest number of links) between any two nodes (institutions) in the network that pass through that edge. If a large number of shortest paths pass through the same edge, then it is in the bottleneck linking communities of nodes. Different colours are used in Figure 3 for different betweenness communities within the three financial markets. Possible contagion paths within communities are little obstructed but such between communities are less accessible. Therefore, contagion will progress differently within the different layers (markets), as they have different betweenness communities.

We provide a detailed comparison in Serguieva (2016, 2017a) of the structure and centralities of single-layers (markets) within any of the available quarterly data periods, and a comparison among periods, concluding decisively that the structures differ. Thus analytical approaches that consider markets are incorporated distinctly (Figures 1 and 3) or indistinctly (Figure 2), within the overall structure of exposures, will observe different contagion processes, identify different systemic risk measures and indexes, and recommend different stabilisations strategies. It is also necessary to evaluate links between markets (see section IV), and then the argument is clear that a multilayer network – incorporating all interconnected markets simultaneously but distinctly – provides the more realistic results.

## III: Formulation and evaluation of single-market contagion dynamics and design of effective stabilisation strategies

### 3.1. Contagion dynamics in the derivatives market

A link in the derivatives market will generally represents how an institution $i$ impacts another institution $j$ in that market – the contribution of $i$ to $j$'s probability of failure – as suggested in (Markose, 2012). We build the structure here involving further details and scenarios and following closely the current regulation and the definition of different exposures data, in comparison with existing studies, and modify the optimisation in approximating the contagion process. First, the probability of failure of an institution $j$ after the start of a contagion process is modelled as dependent on $j$'s own funds and its minimum capital requirement (Serguieva, 2016, 2017a). The contagion dynamics is analysed for the 22 reporting institutions, referred to collectively as 'banks'.

- The current regulatory reporting framework recommended by the Basel Committee on Banking Supervision and implemented in the UK, and the accounting standards with reference to UK GAAP and the International Financial Reporting Standards, look at the different nature of derivatives in comparison with other financial instruments. Banks report their net MtM after collateral derivatives exposures (NAC), and their net derivatives exposures-at-default (EAD). Reported NAC values are non-negative and account for enforceable bilateral netting arrangement[4] between non-defaulted banks throughout

[4] According to the regulatory reporting directives, derivatives transactions are only netted if they are in the same netting set. A 'netting set' is a group of transactions with a single counterparty that are subject to a single, legally enforceable, bilateral netting arrangement. Each transaction that is not subject to a

different netting sets, and for received collateral[5]. The reported exposure-at-default values[6] (EAD) are non-negative and account for collateral, netting arrangement, and adds-on applicable at default (see footnotes 4,5,6), and as a result the EAD amounts are larger than the NAC amounts. We will first use $EAD$ values, and the impact among institutions in the derivatives market will be denoted with the matrix $S^{EAD} = \left[s_{ij}^{EAD}\right]$ of size $n \times n$, where each element $s_{ij}^{EAD}$ reflects a failed bank's $i$ contribution to the default probability of a second bank $j$, and $n$ is the number of reporting institutions. The elements $s_{ij}^{EAD}$ are proportionate to the reported by bank $j$ exposure at default $EAD_{ji}$ to bank $i$, and inversely proportionate to the own funds $C_j$ of bank $j$. The impact matrix $S^{EAD}$ can include both a positive component $s_{ij}^{EAD}$ proportionate to $EAD_{ji}/C_j$ and a positive component $s_{ji}^{EAD}$ proportionate to $EAD_{ij}/C_i$. This is due to received collaterals, netting sets and adds-on applicable at default. In comparison, existing studies on contagion in derivatives markets assume that impact between two institutions exists only in one direction and approximate it as proportionate to the differences in gross exposures.

- Further, when bank $i$ defaults, then the available funds $A_j$ of bank $j$ are reduced with the reported amount of its exposure at default $EAD_{ji}$ to bank $i$. Here, the available funds $A_j = C_j - MC_j$ are the difference between the total own funds $C_j$ and the minimum capital requirements $MC_j$ of $j$. When bank $j$ defaults, the own funds available funds $A_i$ of bank $i$ are reduced with the reported amount of its exposure at default $EAD_{ij}$ to bank $j$. The Own Funds of a bank are evaluated as the sum of its Common Equity Tier 1 Capital (CET1), Additional Tier 1 Capital (AT1), and Tier 2 Capital (T2). The Minimum Capital Requirements to be maintained by a bank are set by current regulation as a percentage of its Total Risk Exposure Amount (risk weighted assets), including buffers in the case of some institutions, and verified by the Prudential Regulatory Authority.

- The non-negative impacts $s_{ij}^{EAD}$ include the case when $j$ receives greater collateral from $i$ that brings the reported exposure to zero (see footnote 5). If bank $j$ does not report an exposure $EAD_{ji}$ to $i$ because the two institutions do not interact in the derivatives market (though they may interact in the fixed-income and/or the securities-financing markets) then $s_{ij}^{EAD} = 0$. When the exposure of $j$ to $i$ is below the reporting threshold then again $s_{ij}^{EAD} = 0$, as $i$ does not significantly impact $j$ directly and so does affect the structural analysis insignificantly.

legally enforceable bilateral netting arrangement is interpreted as its own netting set. Where cross-product netting is legally enforceable, such transactions are considered 'nettable'.

[5] According to the regulatory reporting directives, Net MtM After Collateral for a netting set is computed as Net MtM Before Collateral less the value of collateral received from a counterparty to collateralise the exposure of that netting set. The collateral includes the one received under legally enforceable credit support annexes, as well as any collateral held in excess of what is legally required. The collateral only represents what is received / is in hand on a confirmed settlement basis, and does not include collateral owed to but not actually held by the firm. When collateral received is greater than Net MtM Before Collateral, then Net MtM After Collateral is zero.

[6] Exposure At Default (EAD) is the counterparty credit risk exposure net of collateral, as specified in the Prudential Requirements for Banks, Building Societies and Investment Firms BIPRU 13, and calculated either using the Mark-to-Market Method (BIPRU 13.4), the Standardised Method (BIPRU 13.5), or the Internal Model Method (BIPRU 13.6).

Therefore, the derivatives layer here is built as accurately as possible, using reported data without attempting approximation. In comparison, most studies work with aggregated data and approximate institution-to-institution exposures and impact. However, approximated structures differ from empirical systems in a way that cannot be anticipated, and thus mislead analysis and regulatory implications (Cont et al., 2013).

Here, we consider the boundary case of a single market in isolation, when it is not aware of the liabilities in other markets. An intermediate case is to assume that institutions in the single market are aware of their overall but not bilateral liabilities in other markets, and the approach presented here can also be applied for that case. The intermediate case will account for the overall amount of exposures, but not for the dynamics of activating exposures in other layers and propagating impact among institutions and markets. The case when the multiple market system is aware of all granular exposures is analysed in Section IV.

- When the available funds $A_i$ of $i$ deplete, the bank is considered failed. Therefore, $p_i = \frac{A_i}{C_i}$ is the percentage of own funds that can be used to cover triggered exposures, and $p_i$ differs from institution to institution. Even if two banks $i$ and $j$ have equal total own funds $C_i = C_j$, they may have very different minimum capital requirements $MC_i \neq MC_j$, and therefore different ratios $p_i \neq p_j$. Within the database used here, the ratios $p_i$ differ up to a factor of 4, $\max_{1 \le i,j \le n} (p_i/p_j) \approx 4$. In comparison, existing studies assume that $p$ is the same for all institutions and does not depend on risk weighted assets.

  Assuming $p$ is the same corresponds to a spreading rate $(1-p)$ in the contagion process, for each institution. Instead, we set $p = p_{min} = \min_{1 \le i \le n} (p_i)$, which corresponds to a maximum spreading rate $(1-p_{min})$, but then modify in Equation (1) the condition for default of each bank $i$ at step $(q+1)$ in the contagion process. Instead of using $\sum_{\substack{j \in B_q \\ i \notin B_q}} (s_{ji}^{EAD}) > p$, we use $\sum_{\substack{j \in B_q \\ i \notin B_q}} (s_{ji}^{EAD}) > p_i$:

$$\sum_{\substack{j \in B_q \\ i \notin B_q}} (s_{ji}^{EAD}) = \sum_{\substack{j \in B_q \\ i \notin B_q}} \left(\frac{EAD_{ij}}{C_i}\right) > p_i = \alpha_i \; p_{min} \qquad \text{for } \alpha_i > 1 \qquad (1)$$

  Here, $B_q$ is the set of banks defaulted by step $q$ and $B_q = \cup_{k=1}^{q} \beta_k$, where $\beta_k$ represents the set of banks failed at step $k$. If the own funds of institution $i$ in the denominator are modified to $C_i^{modified} = a_i C_i > C_i$, then the equivalent condition for $i$ defaulting at step $(q+1)$ is:

$$\sum_{\substack{j \in B_q \\ i \notin B_q}} \left(\frac{EAD_{ij}}{C_i^{modified}}\right) = \sum_{\substack{j \in B_q \\ i \notin B_q}} \left(\frac{EAD_{ij}}{\alpha_i C_i}\right) > p_{min} \qquad \text{for } \alpha_i > 1 \qquad (2)$$

  The unique $\alpha_i$ for each bank $i$ is applied, and though the spreading rate is $(1-p_{min})$, the unique default condition for each institution and its unique spreading rate is incorporated into the contagion dynamics through $\alpha_i$.

In order to describe the contagion process, we follow the logic in Firfine (2003) and Markose (2012), and extend with the steps listed above as well as with additional details and clarification:

- **step $q = 0$**
  A set of banks fail at time $q = 0$. This is due to a trigger that is internal or external to the system of reporting banks. It is not known what trigger will be active and which banks will fail. However, if the defaulted banks are denoted with $\beta_0$, then the probability of default of a bank $i \in \beta_0$ at $q = 0$ is assumed as $\underset{i \in \beta_0}{\pi_{i,0}^{EAD}} = 1$.
  The probability of default of the other banks $i \notin \beta_0$ is assumed as insignificantly small $0 < \underset{i \notin \beta_0}{\pi_{i,0}^{EAD}} = \frac{1}{C_i^{modified}} \ll 1$. Due to the failure of banks $\beta_0$, a contagion process starts, and the model derived here will account for any possible set $\beta_0$.

- **step $q = 1$**
  The set of banks that fail at step $q = 1$ is denoted with $\beta_1$. It is not known which banks participate in $\beta_1$, as the elements of $\beta_0$ are not known in advance. A bank $i \in \beta_1$ fails at step $q = 1$, because $\frac{\sum_{j \in \beta_0} \underset{(i \in \beta_1)}{(EAD_{ij})}}{C_i^{modified}} > p_{min}$, and its probability of default at $q = 1$ is $\underset{i \in \beta_1}{\pi_{i,1}^{EAD}} = 1$. On the other hand, the probability of default of banks $i \in \beta_0$ at $q = 1$ is $\underset{i \in \beta_0}{\pi_{i,1}^{EAD}} = 0$, as they already failed at step at $q = 0$.
  Let us denote the set of banks that have failed by step $q = 1$ as $B_1$, then $B_1 = \beta_0 \cup \beta_1$. For completeness, the set of banks that have failed by $q = 0$ can be denoted as $B_0$ where $B_0 = \beta_0$, and therefore $B_1 = B_0 \cup \beta_1$. The probability of default of a bank $i \notin B_1$ surviving at $q = 1$ is $\underset{i \notin B_1}{\pi_{i,1}^{EAD}} = \frac{\sum_{j \in \beta_0} \underset{(i \notin B_1)}{(EAD_{ij})}}{C_i^{modified}} < p_{min}$.
  Here, $\underset{i \notin \beta_0}{\pi_{i,0}^{EAD}} \approx 0$ are not taken into account as these are insignificantly small.

- **step $q = 2$**
  The set of banks that fail at step $q = 2$ is denoted with $\beta_2$, and the set of banks that have failed by step $q = 2$ is denoted with $B_2$, where $B_2 = B_1 \cup \beta_2$. A bank $i \in \beta_2$ (for $i \notin B_1$) fails at step $q = 2$ because the depletion of its available funds exceeds the threshold $\frac{\sum_{j \in B_1} \underset{(i \in \beta_2)}{(EAD_{ij})}}{C_i^{modified}} > p_{min}$, and its probability of default is $\underset{i \in \beta_2}{\pi_{i,2}^{EAD}} = 1$. The probability of default of banks $i \in B_1$ at step $q = 2$ is $\underset{i \in B_1}{\pi_{i,2}^{EAD}} = 0$, as they already failed at step $q = 0$ or $q = 1$. The probability of default of a bank $i \notin B_2$ surviving at $q = 2$ is $\underset{i \notin B_2}{\pi_{i,2}^{EAD}}$. By analogy with the epidemiology literature, $(1 - p_{min})$ is the rate of infection, which in this case is a rate of 'spreading default' or spreading losses. One percent of bank $i$'s capital probably infected at step $q = 1$ has the potential to infect $(1 - p_{min})$ percent of its capital at step $q = 2$. If a bank fails due to infected (lost) capital it also loses up to $(1 - p_{min})$ percent of its capital that hasn't been infected so far. Then these losses will affect other banks at the next step, etc. The percentage of $i$'s capital probably lost at $q = 1$ is $\underset{i \notin B_1}{\pi_{i,1}^{EAD}} = \frac{\sum_{\substack{j \in \beta_0 \\ (i \notin B_1)}} (EAD_{ij})}{C_i^{modified}}$, which depends on $i$'s exposures to banks that failed prior to $q = 1$. This $\underset{i \notin B_1}{\pi_{i,1}^{EAD}}$ is also $i$'s probability of default at $q = 1$ and has the potential to infect or to bring probable losses of $(1 - p_{min})\underset{i \notin B_1}{\pi_{i,1}^{EAD}}$ percent of its capital at $q = 2$. Exposures of $i$ to banks $j \in \beta_1$ that failed at $q = 1$ are lost at $q = 2$, and also contribute to the probability $\underset{i \notin B_2}{\pi_{i,2}^{EAD}}$ of $i$'s default at $q = 2$. Therefore:

$$\underset{i \notin B_2}{\pi_{i,2}^{EAD}} = (1 - p_{min})\, \underset{i \notin B_2}{\pi_{i,1}^{EAD}} + \sum_{\substack{j \in \beta_1 \\ (i \notin B_2)}} \left( \frac{EAD_{ij}}{C_i^{modified}}\, \underset{j \in \beta_1}{\pi_{j,1}^{EAD}} \right) \quad \text{where} \quad \underset{j \in \beta_1}{\pi_{j,1}^{EAD}} = 1$$

It is not known prior to the start of contagion which banks will default at each step, and the probability $\underset{i \notin B_2}{\pi_{i,2}^{EAD}}$ is derived here for any possible $B_0, B_1, B_2$.

- **step $q$**
  The set of banks that fail at step $q$ is denoted with $\beta_q$, and the set of banks that have failed by step $q$ is denoted with $B_q$, where $B_q = B_{q-1} \cup \beta_q$. A bank $i \in \beta_q$ (for $i \notin B_{q-1}$) fails at step $q$ because $\frac{\sum_{\substack{j \in B_{q-1} \\ (i \in \beta_q)}} (EAD_{ij})}{C_i^{modified}} > p_{min}$, and its probability of default at $q$ is:

$$\underset{i \in \beta_q}{\pi_{i,q}^{EAD}} = 1 \tag{3a}$$

  For banks $i \in B_{q-1}$, the probability of default at step $q$ is:

$$\underset{i \in B_{q-1}}{\pi_{i,q}^{EAD}} = 0 \tag{3b}$$

  as they already failed prior to step $q$. The probability of default of banks $i \notin B_q$ surviving at $q$ is:

$$\underset{i \notin B_q}{\pi_{i,q}^{EAD}} = (1 - p_{min})\, \underset{i \notin B_q}{\pi_{i,q-1}^{EAD}} + \sum_{\substack{j \in \beta_{q-1} \\ (i \notin B_q)}} \left[ \left( \frac{EAD_{ij}}{C_i^{modified}} \right) \underset{j \in \beta_{q-1}}{\pi_{j,q-1}^{EAD}} \right] \tag{3c}$$

  for $\underset{j \in \beta_{q-1}}{\pi_{j,q-1}^{EAD}} = 1$ and $\underset{i \notin B_q}{\pi_{i,q-1}^{EAD}} = (1 - p_{min})\, \underset{i \notin B_q}{\pi_{i,q-2}^{EAD}} + \sum_{\substack{j \in \beta_{q-2} \\ (i \notin B_q)}} \left[ \left( \frac{EAD_{ij}}{C_i^{modified}} \right) \underset{j \in \beta_{q-2}}{\pi_{j,q-2}^{EAD}} \right]$.

- **step $q = q_{stop}$**
  The contagion process ends at $q = q_{stop}$ because all remaining banks fail by $q_{stop}$ or because none of the remaining banks fails at $q_{stop}$.

Equations (3a,b,c) present an iteration in the contagion process, and can be summarised into and approximated with the linear system of equations:

$$\Pi_q^{EAD} = \left[ (1 - p_{min}) I + S'^{EAD} \right] \Pi_{q-1}^{EAD} \tag{4a}$$

where $\Pi_q^{EAD}$ is the non-negative probabilities vector of size $n$ :

$$\Pi_q^{EAD} = \left[ \pi_{1,q}^{EAD}, \cdots, \pi_{i,q}^{EAD}, \cdots, \pi_{n,q}^{EAD} \right]' \quad (4b)$$

The impact matrix $S^{EAD}$ at each step $q$ of the contagion process, $0 < q \le q_{stop}$, is:

$$S^{EAD} = \begin{bmatrix} s_{11}^{EAD} \cdots s_{1j}^{EAD} \cdots s_{1n}^{EAD} \\ \vdots \quad \ddots \quad \vdots \quad \iddots \quad \vdots \\ s_{i1}^{EAD} \cdots s_{ij}^{EAD} \cdots s_{in}^{EAD} \\ \vdots \quad \iddots \quad \vdots \quad \ddots \quad \vdots \\ s_{n1}^{EAD} \cdots s_{nj}^{EAD} \cdots s_{nn}^{EAD} \end{bmatrix} \quad \text{with} \quad s_{ij}^{EAD} = \begin{cases} \dfrac{EAD_{ji}}{c_j^{modified}} \ge 0 \ , \ for \ i \ne j \\ 0 \qquad\qquad\quad , \ for \ i = j \end{cases} \quad (4c)$$

At step $q$ , the impact $s_i^{EAD}$ of bank $i$ on institutions in the derivatives market is:

$$s_i^{EAD} = \sum_{j=1}^{n} \left( s_{ij}^{EAD} \right) = \sum_{j=1}^{n} \left( \frac{EAD_{ji}}{c_j^{modified}} \right) > 0 \quad (5)$$

and bank $j$ is affected with $s_j^{EAD}$ by all institutions' activity in this market:

$$s_j^{EAD} = \sum_{i=1}^{n} \left( s_{ij}^{EAD} \right) = \sum_{i=1}^{n} \left( \frac{EAD_{ji}}{c_j^{modified}} \right) > 0 \quad (6)$$

The contagion dynamics throughout steps from $q = 0$ to $q = q_{stop}$ is expressed as the system of equations:

$$\Pi_{q_{stop}}^{EAD} = \left[ (1 - p_{min})\, I + S'^{EAD} \right]^{q_{stop}} \Pi_0^{EAD} \ . \quad (7)$$

## 3.2. Relative systemic-risk indexes and a structural measure of systemic-risk in a single market

Control systems theory (Nise, 2011) tells us that if the maximum Eigenvalue of $\left[ (1 - p_{min})\, I + S'^{EAD} \right]$ is $\lambda_{\left[ (1-p_{min})\, I + S'^{EAD} \right]}^{\max} > 1$ then the contagion process diverges to the destruction of the banking system at some $q = q_{stop}$. If $\lambda_{\left[ (1-p_{min})\, I + S'^{EAD} \right]}^{\max} < 1$ then the system survives and converges to a steady state at some $q = q_{stop}$ . This stability condition can be formulated in terms of the maximum Eigenvalue $\lambda_{S^{EAD}}^{\max}$ of matrix $S^{EAD}$. Using Eigenvalue shifting and considering that the right and left Eigenvectors have the same corresponding maximum Eigenvalue, i.e. $\lambda_{S^{EAD}}^{\max} = \lambda_{S'^{EAD}}^{\max}$ denoted as $\lambda_{EAD}^{\max}$ , produces the stability condition:

$$\lambda_{\left[ (1-p_{min})\, I + S'^{EAD} \right]}^{\max} = (1 - p_{min}) + \lambda_{EAD}^{\max} < 1 \quad \Rightarrow \quad \lambda_{EAD}^{\max} < p_{min} \quad (8)$$

Further, matrix analysis (Chatelin, 2013) asserts that the largest Eigenvalue of a real-valued non-negative matrix is positive and has positive corresponding right and left Eigenvectors, if the matrix is irreducible. Here, $S'^{EAD}$ is real-valued and non-negative but reducible, and its irreducible submatrix can be identified by applying Tarjan's algorithm. This submatrix, denoted with $S_{connected}^{EAD} = \left[ s(ij)_{connected}^{EAD} \right]$, corresponds to the strongly connected subtensor of rank 2 for the derivatives market. It does not include all reporting banks, however banks outside the strongly connected component have incomparably lower potential to influence the system. Therefore, the Eigenpair analysis is performed on the irreducible submatrix $S_{connected}^{EAD}$, and corresponds to the contagion process:

$$\Pi_{q_{stop}}^{EAD} = \left[\left(1 - p_{\substack{min \\ connected}}^{EAD}\right) I + S'^{EAD}_{connected}\right]^q \Pi_0^{EAD} \tag{9}$$

within the strongly connected component with $m_{EAD} \leq n$ participating banks. The largest Eigenvalue of $S_{connected}^{EAD}$ is $\lambda^{\max}_{S_{connected}^{EAD}}$ and we denote $\lambda^{\max}_{S_{connected}^{EAD}} = \lambda^{\max}_{S'^{EAD}_{connected}} = \lambda^{max}_{\substack{EAD \\ connected}}$. Then the stability condition from Equation (8) transforms into:

$$\lambda^{max}_{\substack{EAD \\ connected}} < p_{\substack{min \\ connected}}^{EAD} = \min_{1 \leq i \leq m_{EAD}} \left(\frac{A(i)_{connected}^{EAD}}{C(i)_{connected}^{EAD}}\right) \tag{10}$$

where $p_{\substack{min \\ connected}}^{EAD}$ is evaluated over the $m_{EAD}$ banks. The Eigenvalue satisfies the following inequalities:

$$\lambda^{max}_{\substack{EAD \\ connected}} \leq \left\|S'^{EAD}_{connected}\right\|_\infty = \max_{1 \leq j \leq m_{EAD}} \left(s(j)_{connected}^{EAD}\right) \tag{10a}$$

$$\lambda^{max}_{\substack{EAD \\ connected}} \leq \left\|S_{connected}^{EAD}\right\|_\infty = \max_{1 \leq i \leq m_{EAD}} \left(s(i)_{connected}^{EAD}\right) \tag{10b}$$

and according to Equations (5,6) this leads to:

$$\lambda^{max}_{\substack{EAD \\ connected}} \leq$$

$$\leq \min\left[\max_{1 \leq i \leq m_{EAD}} \left(\sum_{j=1}^{m_{EAD}} \left(\frac{EAD(ji)_{connected}}{C(j)^{EAD}_{\substack{modified \\ connected}}}\right)\right), \max_{1 \leq j \leq m_{EAD}} \left(\sum_{i=1}^{m_{EAD}} \left(\frac{EAD(ji)_{connected}}{C(j)^{EAD}_{\substack{modified \\ connected}}}\right)\right)\right] \tag{11}$$

In other words, the largest Eigenvalue is bounded by the maximum impact of a bank on the strongly-connected derivatives submarket and by the maximum impact caused by that derivatives submarket on a bank.

Notice that Eigenvalue shifting preserves Eigenvectors, and therefore finding the Eigenpair $\left(\lambda^{max}_{\substack{EAD \\ connected}}, v_{\left[\left(1-p_{\substack{min \\ connected}}^{EAD}\right) I + S'^{EAD}_{connected}\right]}\right)$ of matrix $\left[\left(1 - p_{\substack{min \\ connected}}^{EAD}\right) I + S'^{EAD}_{connected}\right]$ that represents the contagion process is equivalent to finding the Eigenpair $\left(\lambda^{max}_{\substack{EAD \\ connected}}, v_{S'^{EAD}_{connected}}\right)$ of $S'^{EAD}_{connected}$. This Eigenpair is generated here through an iterative optimisation as follows:

$$\vartheta_\tau = \frac{\left(S'^{EAD}_{connected}\right)\vartheta_{\tau-1}}{\left\|\left(S'^{EAD}_{connected}\right)\vartheta_{\tau-1}\right\|_\infty} = \frac{\left(S'^{EAD}_{connected}\right)\vartheta_0}{\left\|\left(S'^{EAD}_{connected}\right)^\tau \vartheta_0\right\|_\infty} \qquad \text{for} \quad \tau \geq 1 \tag{12a}$$

including a normalisation with the infinite norm $\left\|\left(S'^{EAD}_{connected}\right)\vartheta_{\tau-1}\right\|_{\infty}$ at each iteration $\tau$, which assures that Equation (11) is satisfied. This Eigenpair $\left(\lambda^{max}_{EAD\atop connected},\ v_{S'^{EAD}_{connected}}\right)$ is produced at convergence $\vartheta_{\tau} = \vartheta_{\tau-1}$, for $\vartheta_{\tau} = \vartheta_{\tau-1} = S'^{EAD}_{connected}\ v_{S'^{EAD}_{connected}} = \lambda^{max}_{EAD\atop connected}\ v_{S'^{EAD}_{connected}}$. Therefore:

$$v_{S'^{EAD}_{connected}} = \left(S'^{EAD}_{connected}\right)^{-1}\vartheta_{\tau} \tag{12b}$$

If the resulting Eigenvector $v_{S'^{EAD}_{connected}}$ is divided by the square of its Euclidean norm $\left(\left\|v_{S'^{EAD}_{connected}}\right\|_2\right)^2$ then:

$$u = \frac{v_{S'^{EAD}_{connected}}}{\left(\left\|v_{S'^{EAD}_{connected}}\right\|_2\right)^2} \quad \text{and} \quad u'\,v_{S'^{EAD}_{connected}} = 1 \tag{13a,b}$$

Here, vector $u$ corresponds to $\lambda^{max}_{EAD\atop connected}$ and to the right Eigenvector $v$ of the transposed $S'^{EAD}_{connected}$ and satisfy Equation (13b). These are the qualities of the right Eigenvector of the impact matrix $S^{EAD}_{connected}$. So the positive vector $u = u_{S^{EAD}_{connected}}$ gives the ranking, according to their systemic impact, of the banks participating in the strongly connected substructure of the derivatives market. The maximum Eigenvalue satisfies:

$$\lambda^{max}_{EAD\atop connected} = u'_{S^{EAD}_{connected}}\ S'^{EAD}_{connected}\ v_{S'^{EAD}_{connected}} = v'_{S'^{EAD}_{connected}}\ S^{EAD}_{connected}\ u_{S^{EAD}_{connected}} \tag{14}$$

and relates to system's stability. In the condition from Equation (10), the difference $\lambda^{max}_{EAD\atop connected} - p^{EAD}_{min\atop connected}$ can be interpreted as the system's distance from structural stability. If $\lambda^{max}_{EAD\atop connected}$ is only slightly larger than $p^{EAD}_{min\atop connected}$ then the system will be eventually destroyed but the contagion process will take long time, and it may be possible to intervene constructively. If $\lambda^{max}_{EAD\atop connected}$ is quite larger than $p^{EAD}_{min\atop connected}$ then the contagion will be more intense, and the system will be destroyed quickly. Therefore, we can formulate the systemic risk emerging in the derivatives market as the structural measure:

$$SR^{EAD}_{risk} = \begin{cases} \lambda^{max}_{EAD\atop connected} - p^{EAD}_{min\atop connected} > 0 & \text{(area of fragility)} \\ 0\,,\ if\ \lambda^{max}_{EAD\atop connected} - p^{EAD}_{min\atop connected} < 0 & \text{(area of resilience)} \end{cases} \tag{15a}$$

This measure allows comparing the stability of two structures (markets) irrespectively of monetary values. For example, the banking systems in two countries may be similarly instable but involving different monetary values. The objective here, through designing stabilisation strategies in the next Sections, is to build in structural resilience. Then the system will better sustain its associated monetary values.

We also formulate with Equation (15b) the systemic risk index of a bank $i$ in percentages. This can be interpreted as the percentage that $i$ contributes to systemic instability or to the systemic risk $SR_{risk}^{EAD}$ of that market:

$$SRI(i)_{index}^{EAD} = \begin{cases} SRI(i)_{\substack{index\\connected}}^{EAD} = \dfrac{u(i)_{S_{connected}^{EAD}}}{\sum_{i=1}^{m_{EAD}} \left( u(i)_{S_{connected}^{EAD}} \right)} > 0; \quad \text{for} \quad i \in \{1, \cdots, m_{EAD}\} \\ SRI(i)_{\substack{index\\unconnected}}^{EAD} = 0 \;; \quad \text{for} \quad i \in \{m_{EAD}+1, \cdots, n\} \end{cases} \tag{15b}$$

Banks participating in the strongly connected substructure of the market have positive indexes, while banks outside it have zero indexes and do not contribute to the $SR_{risk}^{EAD}$. Here, $SRI(i)_{\substack{index\\connected}}^{EAD}$ are relative measures and $SR_{risk}^{EAD}$ is an absolute measure, due to interconnectivity in the derivatives market. The index $SRI(i)_{\substack{index\\connected}}^{EAD}$ of bank $i$ can be translated in absolute terms as the part $\left( SRI(i)_{\substack{index\\connected}}^{EAD} \; SR_{risk}^{EAD} \right)$ that $i$ contributes to the structural systemic risk $SR_{risk}^{EAD}$.

## 3.3. Stabilisation strategies in a single market

Most studies analysing the structure of financial systems do not quantify systemic risk. The few studies quantifying risk rarely comment on single-layer stabilisation strategies, and multilayer strategies have not been addressed. Existing studies of the derivatives market recommend that capital surcharges are collected only from very few top-ranked systemically important institutions, and set aside in a fund that then can be accessed by any institution when in distress. Such step will be helpful but not optimal. It will not really build in structural resilience into the system, and is not pre-emptive as it expects institutions to fall in distress. When institutions fall in distress, they will need large funding to be able to recover, and such approach is still at a significant cost for the system. The fund may deplete while helping some institutions and not others, as well. We consider that it is not sufficient to collect surcharges but it is important to distribute them optimally among all institutions, and it is necessary to collect them in an optimal cost-effective way. In order to achieve structural balance, not only the very top few institutions should participate in the stabilisation strategy but all institutions with nonzero systemic impact (nonzero systemic risk index). The most important institutions can be viewed as and are 'most guilty', but system's instability is not entirely their fault – it is rather a fault of the emerged structure. Therefore, if a stabilisation strategy subtly and adaptively affects the emergence process of structure, it will build in systemic resilience and achieve pre-emptive stabilisation at a minimum cost. The participation of institutions in the strategy is proportionate to their systemic indexes but with a very small fraction of their capital, and these fractions are immediately redistributed optimally and granularly among the same institutions. The strategy is at no cost for the system, the surcharges are optimised at their minimum for an institution in comparison with other mechanisms, and the participation of any institution is less than its surcharges as it immediately proportionate compensations. The strategy includes a stabilisation step in the current period only if the systemic risk or resilience at the end of the last period was less than a targeted threshold. Therefore, the structure is maintained around the threshold, only minimum adjustments are required, and in some periods they may not be required. This could be implemented as part of the infrastructure mechanism, and would also play the role of monitoring systemic stability. If we look for an analogy, this mechanism may resemble the varying margin within the current clearance mechanisms.

Based on the indexes from Equation (15b), a systemic risk surcharge for an institution $i$ is formulated as:

$$SRS(i)^{EAD}_{surcharge} = \gamma_{EAD}\, SRI(i)^{EAD}_{index} =$$

$$= \begin{cases} SRS(i)^{EAD}_{\substack{surcharge\\connected}} = \gamma_{EAD}\, SRI(i)^{EAD}_{\substack{index\\connected}} & \text{for } \; 0 < \gamma_{EAD} \ll 1\,;\; i \in \{1, \cdots, m_{EAD}\} \\ SRS(i)^{EAD}_{\substack{surcharge\\unconnected}} = 0 & \text{for } \; i \in \{m_{EAD}+1, \cdots, n\} \end{cases} \tag{16}$$

It is applied to evaluate a fraction $\gamma_{EAD}\, SRI(i)^{EAD}_{\substack{index\\connected}}\; C(i)^{EAD}_{\substack{modified\\connected}}$ of its capital. Here, $\gamma_{EAD}$ is very small and optimised to estimate minimum surcharges for each institution $i$ that when distributed in a balancing way to each institution $j$, in proportion to the impact of $i$ on $j$, will bring the system to the targeted structural threshold. This is equivalent to building in structural resilience. The proportion is the ratio of the impact $s(ij)^{EAD}_{connected}$ of bank $i$ on $j$ divided by the overall impact of bank $i$ on the derivatives market, $s(i)^{EAD}_{connected} = \sum_{j=1}^{m_{EAD}} \left( s(ij)^{EAD}_{connected} \right)$, for $i, j \in \{1, \cdots, m_{EAD}\}$. Let us denote with $X(ij)$ the proportion of the surcharge on $i$ distributed to $j$. Equation (17) shows how impact matrix $S^{EAD}_{connected} = \left[ s(ij)^{EAD}_{connected} \right]$ changes into $S^{rebalanced}_{\substack{EAD\\connected}}$:

$$\left[ s(ij)^{rebalanced}_{\substack{EAD\\connected}} \right] = \left[ S(ij)_{\substack{EAD\\connected}} \Bigg/ \left( 1 + \sum_{i=1}^{m_{EAD}} \left( \frac{X(ij)}{p^{EAD}_{\substack{min\\connected}}\; C(j)^{EAD}_{\substack{modified\\connected}}} \right) \right) \right] =$$

$$= \left[ \frac{S(ij)_{\substack{EAD\\connected}}}{1 + \sum_{i=1}^{m_{EAD}} \left( \gamma_{EAD}\; SRI(i)^{EAD}_{\substack{index\\connected}} \left( \frac{C(i)^{EAD}_{\substack{modified\\connected}}}{p^{EAD}_{\substack{min\\connected}}\; C(j)^{EAD}_{\substack{modified\\connected}}} \right) \left( \frac{S(ij)_{\substack{EAD\\connected}}}{\sum_{q=1}^{m_{EAD}} \left( S(iq)_{\substack{EAD\\connected}} \right)} \right) \right)} \right] \tag{17}$$

It considers that the funds $A(j)^{EAD}_{connected} = C(j)^{EAD}_{connected} - MC(j)^{EAD}_{connected}$ available to $j$ increase to $A(j)^{\substack{EAD\\rebalanced}}_{connected} = A(j)^{EAD}_{connected} + \sum_{i=1}^{m_{EAD}} X(ij)$ with the proportionate fractions $X(ij)$. In Section 3.1, we denoted the ratio of available to total own funds of $j$ as $p(j)^{EAD}_{connected} = a(j)^{EAD}_{connected}\ p^{EAD}_{\substack{min\\connected}} = \frac{A(j)^{EAD}_{connected}}{C(j)^{EAD}_{connected}} = \frac{a(j)^{EAD}_{connected} A(j)^{EAD}_{connected}}{C(j)^{EAD}_{\substack{modified\\connected}}}$

Maintaining the parameter $p^{EAD}_{\substack{min\\connected}} = p^{\substack{EAD\\rebalanced}}_{\substack{min\\connected}}$ in the simulation of contagion within the rebalanced structure leads to:

$$p^{EAD}_{\substack{min\\connected}} = \frac{A(j)^{EAD}_{connected}}{C(j)^{EAD}_{\substack{modified\\connected}}} = \frac{A(j)^{\substack{EAD\\rebalanced}}_{connected}}{C(j)^{\substack{EAD\\rebalanced}}_{\substack{modified\\connected}}} = \frac{A(j)^{EAD}_{connected} + \sum_{i=1}^{m_{EAD}} X(ij)}{C(j)^{\substack{EAD\\rebalanced}}_{\substack{modified\\connected}}} \tag{18a}$$

and to a new modified value $C(j)^{\substack{EAD\\rebalanced}}_{\substack{modified\\connected}}$ after rebalancing:

$$C(j)^{\substack{EAD\\rebalanced}}_{\substack{modified\\connected}} = C(j)^{EAD}_{\substack{modified\\connected}} \left(1 + \frac{\sum_{i=1}^{m_{EAD}} X(ij)}{p^{EAD}_{\substack{min\\connected}}\ C(j)^{EAD}_{\substack{modified\\connected}}}\right) \tag{18b}$$

This produces the denominator in Equation (17), because $s(ij)_{\substack{EAD\\connected}} = \frac{EAD(ji)}{C(j)^{EAD}_{\substack{modified\\connected}}}$

and:

$$s(ij)^{rebalanced}_{\substack{EAD\\connected}} = \frac{EAD(ji)}{C(j)^{\substack{EAD\\rebalanced}}_{\substack{modified\\connected}}} =$$

$$= \frac{EAD(ji)}{C(j)^{EAD}_{\substack{modified\\connected}} \left(1 + \frac{\sum_{i=1}^{m_{EAD}} X(ij)}{p^{EAD}_{\substack{min\\connected}}\ C(j)^{EAD}_{\substack{modified\\connected}}}\right)} = \frac{s(ij)_{\substack{EAD\\connected}}}{\left(1 + \frac{\sum_{i=1}^{m_{EAD}} X(ij)}{p^{EAD}_{\substack{min\\connected}}\ C(j)^{EAD}_{\substack{modified\\connected}}}\right)} \tag{18c}$$

The rebalancing preserves $S^{rebalanced}_{\substack{EAD\\connected}}$ as non-negative, and the Eigenpair analysis can be validly applied. Equation (17) reduces $\max_{1 \le i \le m_{EAD}} \left(s(j)^{rebalanced}_{\substack{EAD\\connected}}\right)$ and $\max_{1 \le j \le m_{EAD}} \left(s(i)^{rebalanced}_{\substack{EAD\\connected}}\right)$, and from Equation (11) it follows that:

$$\lambda^{EAD_max}_{\substack{rebalanced\\connected}} \le \min\left[\max_{1 \le i \le m_{EAD}} \left(s(j)^{rebalanced}_{\substack{EAD\\connected}}\right), \max_{1 \le j \le m_{EAD}} \left(s(i)^{rebalanced}_{\substack{EAD\\connected}}\right)\right] \le \lambda^{EAD_max}_{connected} \tag{19}$$

The largest Eigenvalue is reduced[7], which is equivalent to increasing structural resilience. The parameter $\gamma_{EAD}$ is identified, through search and optimisation, as the smallest value that when applied in Equation (17) transforms the system $S_{\substack{EAD\\connected}}$ into a system $S^{rebalanced}_{\substack{EAD\\connected}}$ with targeted threshold $SR^{EAD}_{\substack{risk\\threshold}}$. With a minimum structural change, the value of systemic risk in Equation (15a) moves in direction towards the area of resilience.

[7] If the model is considered without the financial context, then reducing the maximum Eigenvalue can be attempted alternatively. For example, by reducing the sum of elements in a row of the transposed $\left[s(ji)_{\substack{EAD\\connected}}\right]'$ by increasing the denominator of the elements with a factor of $(1 + \delta)$. (notice that each element in a row is $s(ji)_{\substack{EAD\\connected}} = EAD(ij)_{connected} \big/ C(i)^{EAD}_{\substack{modified\\connected}}$ and has the same

| Empirical Contagion Dynamics in the Derivatives Market<br>based on data for one of the quarters in the period from June 2014 to September 2015 | Table 1 |
|---|---|
| number of reporting banks | $n = 22$ |
| number of banks in the strongly connected subtensor | $m_{EAD} = 19$ |
| $p^{EAD}_{\substack{min\\connected}}$ | 0.14573 |
| stability condition | $\lambda^{max}_{\substack{EAD\\connected}} < 0.14573$ |
| $\lambda^{max}_{\substack{EAD\\connected}}$ for $\gamma_{EAD} = 0$ | 0.07268 |
| $SR^{EAD}_{resilience}$ for $\gamma_{EAD} = 0$ | 0.07305 |

The empirical analysis next is performed for one of the quarters in the period from June 2014 to September 2015, and the results are presented in Table 1. In that quarter, 19 out of the 22 reporting institutions participate in the strongly connected component within the structure emerging from interlinkages in the derivatives market. Therefore, 19 institutions have nonzero systemic-risk indexes and affect structural stability. The largest Eigenvalue is 0.07268 and satisfies the condition $\lambda^{max}_{\substack{EAD\\connected}} < 0.14573$, indicating that the system is in the area of structural resilience. We can define a measure $SR^{EAD}_{resilience}$ of structural resilience as:

$$SR^{EAD}_{resilience} = \begin{cases} 0\,,\ if\ p^{EAD}_{\substack{min\\connected}} - \lambda^{max}_{\substack{EAD\\connected}} < 0 \\ p^{EAD}_{\substack{min\\connected}} - \lambda^{max}_{\substack{EAD\\connected}} > 0 \end{cases} \tag{20}$$

If $p^{EAD}_{\substack{min\\connected}}$ is only slightly larger than $\lambda^{max}_{\substack{EAD\\connected}}$, the contagion process will eventually be contained but this will take long time, and a number of institutions will default though part of the system will survive. If $p^{EAD}_{\substack{min\\connected}}$ is quite larger than denominator) In the financial context here, this will mean that we charge an institution $i$ with a fraction of its capital and then use that fraction to increase the capital of the same institution. The meaning of a systemic risk charge for $i$, however, is rather to increase funds available to institutions affected by $i$ and so reduce the impact of $i$ on them.

$\lambda^{max}_{\substack{EAD\\connected}}$, then the contagion will be contained quickly and a large part of the system will survive. The empirical result here is $SR^{EAD}_{resilience} = 0.07305$. For a threshold of $SR^{EAD}_{\substack{risk\\threshold}} = 0$, no stabilisation step is necessary at the start of the next quarterly period, and therefore results of simulating stabilisation strategies are not included in Table 1. We will note, however, that any movement in direction towards smaller $SR^{EAD}_{risk} > 0$ or larger $SR^{EAD}_{resilience} > 0$ is equivalent to building in resilience. For example, a meta strategy may involve different thresholds $SR^{EAD}_{\substack{risk\\threshold}}(t_k) > 0$ in different periods $t_k$ , $1 \leq k \leq T$, so that the system gradually moves to a long-term target. A meta strategy may also involve buffer thresholds $SR^{EAD}_{\substack{resilience\\threshold}}(t_k) > 0$ in some periods, as the current contagion and stabilisation analysis is in response to a trigger and the contagion it activates, but does not account for two different triggers activating a second contagion processes while the first is still running or just after it ends. A threshold must be selected carefully for a subtle effect, and the selection may depend on the scope, size and monetary value of the system or subsystem being analysed.

## Comparative Empirical Results under NAC and EAD scenarios

based on data for one of the quarters in the period from June 2014 to September 2015

Table 2

| NAC | | EAD | |
|---|---|---|---|
| number of reporting banks | $n = 22$ | $n = 22$ | number of reporting banks |
| number of banks in the strongly connected subtensor | $m_{NAC} = 16$ | $m_{EAD} = 19$ | number of banks in the strongly connected subtensor |
| $p^{NAC}_{\substack{min\\connected}}$ | 0.26843 | 0.14573 | $p^{EAD}_{\substack{min\\connected}}$ |
| stability condition | $\lambda^{max}_{\substack{NAC\\connected}} < 0.26843$ | $\lambda^{max}_{\substack{EAD\\connected}} < 0.14573$ | stability condition |
| $\lambda^{max}_{\substack{NAC\\connected}}$ for $\gamma_{NAC} = 0$ (no rebalance implemented) | 0.00715 | 0.07268 | $\lambda^{max}_{\substack{EAD\\connected}}$ for $\gamma_{EAD} = 0$ (no rebalance implemented) |
| $SR^{NAC}_{resilience}$ for $\gamma_{NAC} = 0$ | 0.26128 | 0.07305 | $SR^{EAD}_{resilience}$ for $\gamma_{EAD} = 0$ |

Notice that Equations (4a,7) represent a more intensive contagion dynamics (a boundary scenario) than Equations (3a,b,c). The formulation of $\left[s(ij)^{EAD}_{connected}\right]$ corresponds to analysis of a structure functioning as if the going-concern exposures to non-failed banks were also equal to the exposures at-default. The going concern principle in accounting is the assumption that an entity will remain in business for the foreseeable future. Next, we will perform the analysis of the derivatives layer for a structure functioning as if the going-concern exposures are equal to the net MtM exposures after collateral (NAC). These are the correct going-concern exposures, because up until its failure, a non-failed bank $i$ affects with NAC exposures the other non-failed banks $j$. The NAC-scenario is also boundary, as it assumes that a failed bank $i$ affects with the going-concern exposure NAC a non-failed bank $j$, instead with the exposure at-default EAD . The reported non-negative $NAC_{ij}$ , for $1 \le i,j \le n$ , account for received collateral and for enforceable bilateral netting arrangement between non-defaulted banks throughout different netting sets. The tensor (structure) can include both a positive impact $s_{ij}^{NAC} > 0$ of bank $i$ on bank $j$ proportionate to $NAC_{ji}$ , and a positive impact $s_{ij}^{NAC} > 0$ of bank $j$ on bank $i$ proportionate to $NAC_{ij}$. Next, the steps described above for the $S^{EAD}$ analysis are now applied to $S^{NAC}$, and lead to evaluating the Eigenpair $\left(\lambda^{max}_{\substack{NAC\\connected}} , u_{S^{NAC}_{connected}}\right)$ of the strongly connected substructure $S^{NAC}_{connected} = \left[s(ij)^{NAC}_{connected}\right]$, the indexes:

$$SRI(i)^{NAC}_{\substack{index\\connected}} = \frac{u(i)_{S^{NAC}_{connected}}}{\sum_{i=1}^{m_{NAC}}\left(u(i)_{S^{NAC}_{connected}}\right)} \qquad \text{for} \quad i \in \{1, \cdots, m_{NAC}\} \tag{21a}$$

and the resilience:

$$SR^{NAC}_{resilience} = \begin{cases} 0\,,\ if\ \ \lambda^{max}_{\substack{NAC\\connected}} - p^{NAC}_{\substack{min\\connected}} \ge 0 \\ \left|\lambda^{max}_{\substack{NAC\\connected}} - p^{NAC}_{\substack{min\\connected}}\right| \ ,\ if\ \lambda^{max}_{\substack{NAC\\connected}} - p^{NAC}_{\substack{min\\connected}} < 0 \end{cases} \tag{21b}$$

Systemic Risk Ranking and Indexes in the Derivatives Market

based on data for one of the quarters in the period from June 2014 to September 2015

Table 3

| institutions | A | B | C | D |
|---|---|---|---|---|
| rank at $\gamma_{NAC} = 0$, (going-concern systemic dynamics) | 10 | 8 | 0 (not participating in the fragility strongly-connected component) | 7 |
| rank at $\gamma_{EAD} = 0$, (at-default systemic dynamics) | 3 | 8 | 10 | 15 |
| $SRI^{NAC}_{\substack{index\\connected}}(i)$ at $\gamma_{NAC} = 0$ | 2.34% | 3.01% | 0% | 4.00% |
| $SRI^{EAD}_{\substack{index\\connected}}(i)$ at $\gamma_{EAD} = 0$ | 13.15% | 5.09% | 4.03% | 0.81% |

Tables 2 and 3 report and compare empirical results for the NAC-scenario and the EAD-scenario. The structural resilience of the empirical system under the NAC-scenario is $SRR^{NAC}_{resilience} = 0.26128$ , which is higher than the resilience under the EAD-scenario $SRR^{EAD}_{resilience} = 0.07305$. Different number of reporting banks have nonzero structural impact, $m_{NAC} = 16 \neq m_{EAD} = 19$, and participate in the corresponding two strongly connected components. The ranking and index of each bank are different under the two scenarios. The institution encoded with $A$ in Table 3 is of higher ranking under EAD but lower ranking under NAC, which is also confirmed by its corresponding indexes. The opposite is true for institution $D$, it is of higher ranking under NAC and of lower ranking under EAD. Institution $B$ has the same rank 8 among the $m_{NAC}$ banks and among the $m_{EAD}$ banks, but it has different indexes $SRI^{NAC}_{\substack{index \\ connected}}(B) = 3.01\%$ and $SRI^{EAD}_{\substack{index \\ connected}}(B) = 5.09\%$. Bank $C$ is of medium ranking under EAD and is not ranked under NAC, therefore has zero structural impact $SRI^{NAC}_{\substack{index \\ connected}}(C) = 0$.

The empirical results confirm that if we would like to introduce subtle changes in the structure in order to increase its resilience, then different banks and to a different extent will participate in a strategy under each of the two scenarios. NAC and EAD are boundary scenarios, and the strategy can be formulated with surcharges depending both on NAC and EAD indexes, instead. In the terminology, we will use from now on 'systemic-impact index' $SII(i)$ instead of systemic-risk index $SRI(i)$, and correspondingly 'systemic-impact surcharge' $SIS(i)$ instead of systemic-risk surcharge $SRS(i)$. This terminology accounts for the fact that the index measures the proportionate contribution of an institution to systemic risk, but also for the fact that this potential of an institution for structural impact can be used in stabilisation strategies to build in structural resilience. In the case of the EAD and NAC scenarios, the new terminology translates as:

$$SIS(i)_{surcharge} = f\left( \gamma_{EAD}\, SII(i)^{NAC}_{\substack{index \\ connected}},\ \gamma_{NAC}\, SII(i)^{EAD}_{\substack{index \\ connected}} \right) \tag{22}$$

In comparison, Poledna et al. (2015) and Markose (2012) do not differentiate between the two types of derivatives exposure. The contagion algorithm in Poledna et al. (2015) prevents a failed bank to have effect beyond the period of its failure. The approach presented here builds in targeted resilience even when none of the institutions fails. It also does not directly restrict and so preserves the emerged preferences of interaction among banks, and so introduces minimum changes to the system. However, it introduces an incentive for institutions to adapt their preferences to the emergence of a more resilient structure of interactions. A next task is to extend the algorithm to provide that the effect of a non-failed bank is proportionate to NAC exposures, the effect of a failed bank is proportionate to EAD exposures, and a failed bank has no effect beyond the period it fails.

## IV: Formulation and evaluation of multiple-market contagion dynamics and stabilisation strategies

Banks interact simultaneously in multiple markets. The database used here accounts for the interaction of reporting institutions in the fixed-income market, securities-financing market and derivatives market. Section III above does not consider simultaneously contagion dynamics due to connectivity within all markets and among markets. If a bank $i$ is highly affected in the fixed income market by failing banks $j \epsilon \{1, \dots, n\}$, then the position of bank $i$ in the derivatives market is affected by the probability of $i$ failing due to its interaction in the fixed-income market. In other words, the interaction of bank $i$ within the fixed-income market has an impact on its interaction within the derivatives market, and contributes to the probability of bank $i$ failing due to interlinkages in the derivatives market.

## 4.1. Theoretical formulation

A model incorporating simultaneously but distinctly all interconnected markets can be formulated as a tensor-multiplex (Serguieva, 2016, 2017a), where $S$ is a tensor of rank four:

$$S = \sum_{k=1}^{m} \sum_{\ell=1}^{m} \sum_{j=1}^{n} \sum_{i=1}^{n} \left( S^{i\ \ell}_{\ j\ k} \right) \vec{\varepsilon}_i \otimes \vec{\omega}^{j\,\prime} \otimes \vec{\varepsilon}_\ell \otimes \vec{\omega}^{k\,\prime} \tag{23}$$

$$\text{for } \ S^{i\ \ell}_{\ j\ k} \begin{cases} = 0 & \text{if} \quad i = j \wedge \ell = k \vee i \neq j \wedge \ell \neq k \\ \geq 0 & \text{if} \quad i \neq j \wedge \ell = k \vee i = j \wedge \ell \neq k \end{cases}$$

where $m = 3$ corresponds to the three markets, i.e. $k, \ell = 1$ for the fixed-income market, $k, \ell = 2$ for the securities-financing market, and $k, \ell = 3$ for the derivatives market. The number of institutions is $n$, and $S^{i\ \ell}_{\ j\ k} \geq 0$ is the impact of bank $i$ – due to its interaction in market $\ell$ – on institution $j$ acting in market $k$. The impact $S^{i\ \ell}_{\ j\ k} \geq 0$ between two different institutions $i \neq j$ is due to their interaction within the same market $\ell = k$, while the impact is $S^{i\ \ell}_{\ j\ k} = 0$ when we consider $i$ and $j$ as acting in different markets $\ell \neq k$. Further, $S^{i\ \ell}_{\ j\ k} \geq 0$ when the same institution $i = j$ acts in different markets $\ell \neq k$, while $S^{i\ \ell}_{\ j\ k} = 0$ when this institution acts in the same market $\ell = k$. An interconnected multiplex is a multilayer network where mostly the same nodes participate in different type of interactions (interdependencies), and the interaction of a node due to one type of activities is dependent on its interaction due to another type of activities. A tensor can be considered as an interconnected multiplex that also incorporates a basis (innate) structure. In Equation (23), $\vec{\varepsilon}_i \otimes \vec{\omega}^{j\,\prime} \otimes \vec{\varepsilon}_\ell \otimes \vec{\omega}^{k\,\prime}$ stands for the basis structure that includes four vectors $\vec{\varepsilon}_i$, $\vec{\omega}^{j\,\prime}$, $\vec{\varepsilon}_\ell$, $\vec{\omega}^{k\,\prime}$ in their cohesion or tensor multiplication, hence the tensor is of rank four. These vectors characterise, correspondingly, institutions $i$, institutions $j$, markets $\ell$, and markets $k$, for $i, j \in \{1, \cdots, n\}$ and $\ell, k \in \{1, \cdots, m\}$. Tensor-multiplex models expand the scope of feasible structural analysis and stress testing of the financial system (Serguieva, 2016, 2017a). Here, they are only used in modelling contagion and stabilisation processes within multiple interconnected markets.

We build the tensor of rank four as including nine subtensors of rank 2 (see Figure 4). The impact matrix $\left[s(ij)_D\right]$ in the derivatives market $(D)$ has the same meaning as $\left[s_{ij}^{EAD}\right]$ in Section III and:

$$S^{i\ \ell}_{\ j\ k} = \begin{cases} \left( S^{i\ \ \ell=D}_{\ j\neq i\ k=D} \right) = s(ij)_D = s(ij)_{EAD} \\ \left( S^{i\ \ \ell=D}_{\ j=i\ k=D} \right) = s(ii)_D = 0 \end{cases} \quad \text{for} \quad i, j \epsilon \{1, \cdots, n\} \tag{24a}$$

Banks report to the PRA database their exposures in the fixed-income market $(FI)$ as gross MtM values, then $MtM(ji)_{FI}$ will denote the exposure of bank $j$ to bank $i$ in the $FI$ layer. Banks also report their exposures in the securities-financing market $(SF)$ as gross Notional values, then $Notional(ji)_{SF}$ will denote the exposure of institution $j$ to institution $i$ in the $SF$ layer. This reported information does not allow differentiating between going-concern and at-default multiplex exposures.

**Figure 4:** *Four-dimensional structure of impact*
It captures impact among institutions within each financial market and between any pair of markets.

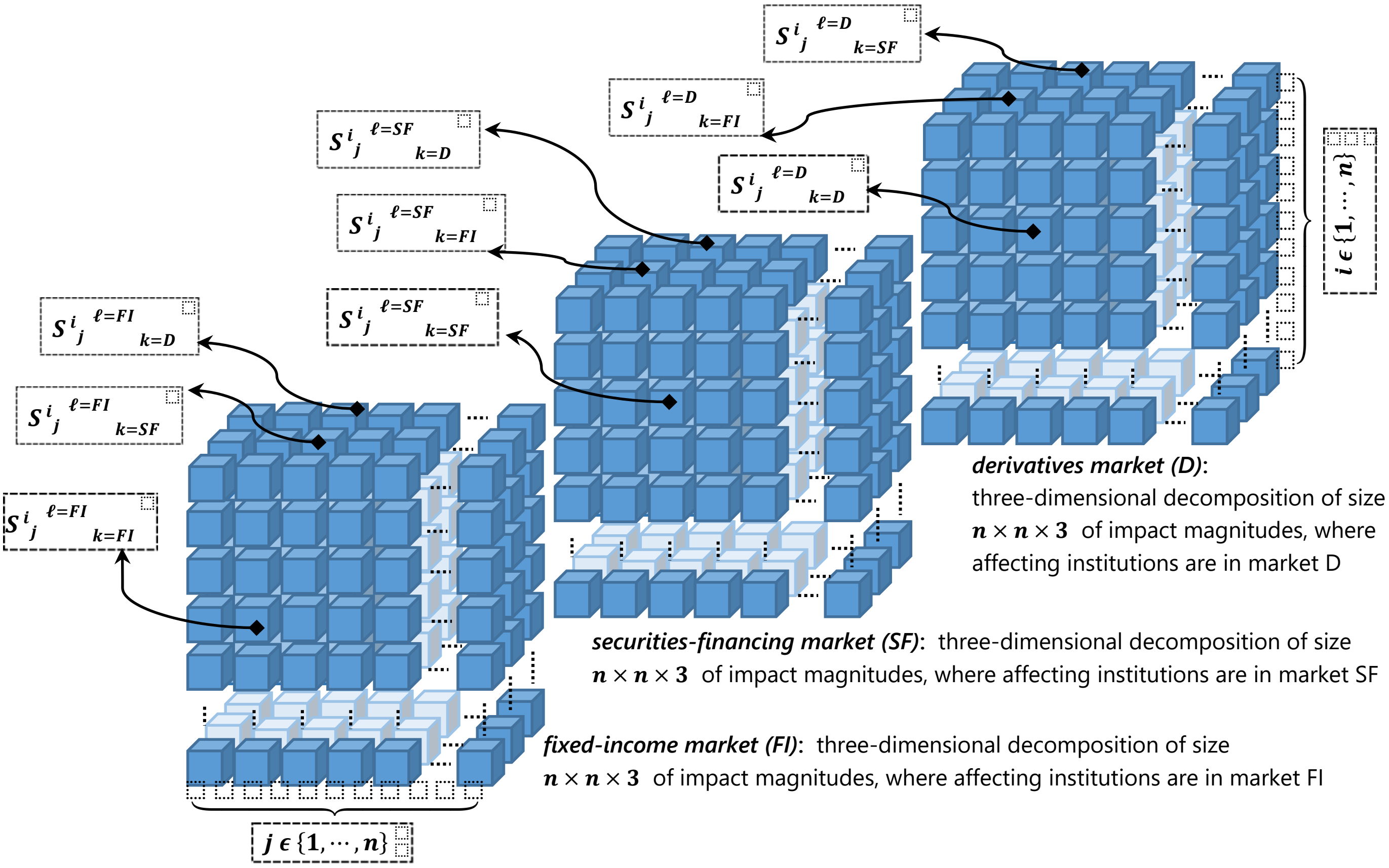

The impact structure $\left[S^{i}{}_{j}{}^{\ell}{}_{k}\right]^{multiplex}$ will be evaluated as follows. The impact matrix $[s(ij)_{FI}]$, due to interconnectivity in the $FI$ market, has elements:

$$S^{i}{}_{j}{}^{\ell}{}_{k} = \begin{cases} \left(S^{i}{}_{j\neq i}{}^{\ell=FI}{}_{k=FI}\right) = s(ij)_{FI} = \begin{cases} \frac{MtM(ji)_{FI} - MtM(ij)_{FI}}{C(j)} > 0 \\ 0\,, if\ \frac{MtM(ji)_{FI} - MtM(ij)_{FI}}{C(j)} \leq 0 \end{cases} \\ \left(S^{i}{}_{j=i}{}^{\ell=FI}{}_{k=FI}\right) = s(ii)_{FI} = 0 \qquad \text{for } i, j \epsilon \{1, \cdots, n\} \end{cases} \tag{24b}$$

where $C(j)$ are the total own funds of bank $j$. The impact matrix $[s(ij)_{SF}]$, due to interconnectivity in the $SF$ market, has elements:

$$S^{i}{}_{j}{}^{\ell}{}_{k} = \begin{cases} \left(S^{i}{}_{j\neq i}{}^{\ell=SF}{}_{k=SF}\right) = s(ij)_{SF} = \begin{cases} \frac{Notional(ji)_{SF} - Notional(ij)_{SF}}{C(j)} > 0 \\ 0\,, if\ \frac{Notional(ji)_{SF} - Notional(ij)_{SF}}{C(j)} \leq 0 \end{cases} \\ \left(S^{i}{}_{j=i}{}^{\ell=SFT}{}_{k=SF}\right) = s(ii)_{SF} = 0 \qquad \text{for } i, j \epsilon \{1, \cdots, n\} \end{cases} \tag{24c}$$

The impact magnitudes between markets are correspondingly:

- $\left[s(ij)_{FI\to SF}\right]$ composed by the impact of institutions $i$ in the fixed-income market $\ell = FI$ on institutions $j$ in the securities-financing market $k = SF$ :

$$S^{i\ \ell}_{\ j\ k} = \begin{cases} \left(S^{i\ \ \ell=FI}_{\ j=i\ \ k=SF}\right) = s(ii)_{FI\to SF} = \sum_{q=1}^{n} s(qi)_{FI} \\ \left(S^{i\ \ \ell=FI}_{\ j\neq i\ \ k=SF}\right) = s(ij)_{FI\to SF} = 0 \end{cases} \quad \text{for } i,j,q\epsilon\{1,\cdots,n\} \qquad (25a)$$

- $\left[s(ij)_{FI\to D}\right]$ composed by the impact of banks $i$ in market $\ell = FI$ on banks $j$ in market $k = D$ :

$$S^{i\ \ell}_{\ j\ k} = \begin{cases} \left(S^{i\ \ \ell=FI}_{\ j=i\ \ k=D}\right) = s(ii)_{FI\to D} = \sum_{q=1}^{n} s(qi)_{FI} \\ \left(S^{i\ \ \ell=FI}_{\ j\neq i\ \ k=D}\right) = s(ij)_{FI\to D} = 0 \end{cases} \quad \text{for } i,j,q\epsilon\{1,\cdots,n\} \qquad (25b)$$

- $\left[s(ij)_{SF\to FI}\right]$ comprises the impact of $i$ in market $\ell = SF$ on $j$ in market $k = FI$ , and $\left[s(ij)_{SF\to D}\right]$ comprises the impact of $i$ in market $\ell = SF$ on $j$ in market $k = D$ :

$$S^{i\ \ell}_{\ j\ k} = \begin{cases} \left(S^{i\ \ \ell=SF}_{\ j=i\ \ k=FI}\right) = s(ii)_{SF\to FI} = \sum_{q=1}^{n} s(qi)_{SF} \\ \left(S^{i\ \ \ell=SF}_{\ j\neq i\ \ k=FI}\right) = s(ij)_{SF\to FI} = 0 \end{cases} \quad \text{for } i,j,q\epsilon\{1,\cdots,n\} \qquad (26a)$$

$$S^{i\ \ell}_{\ j\ k} = \begin{cases} \left(S^{i\ \ \ell=SF}_{\ j=i\ \ k=D}\right) = s(ii)_{SF\to D} = \sum_{q=1}^{n} s(qi)_{SF} \\ \left(S^{i\ \ \ell=SF}_{\ j\neq i\ \ k=D}\right) = s(ij)_{SF\to D} = 0 \end{cases} \quad \text{for } i,j,q\epsilon\{1,\cdots,n\} \qquad (26b)$$

- $\left[s(ij)_{D\to FI}\right]$ includes the impact of $i$ in market $\ell = D$ on $j$ in market $k = FI$ , and $\left[s(ij)_{D\to SF}\right]$ includes the impact of $i$ in market $\ell = D$ on $j$ in market $k = SF$ :

$$S^{i\ \ell}_{\ j\ k} = \begin{cases} \left(S^{i\ \ \ell=D}_{\ j=i\ \ k=FI}\right) = s(ii)_{D\to FI} = \sum_{q=1}^{n} s(qi)_{D} \\ \left(S^{i\ \ \ell=D}_{\ j\neq i\ \ k=FI}\right) = s(ij)_{D\to FI} = 0 \end{cases} \quad \text{for } i,j,q\epsilon\{1,\cdots,n\} \qquad (27a)$$

$$S^{i\ \ell}_{\ j\ k} = \begin{cases} \left(S^{i\ \ \ell=D}_{\ j=i\ \ k=SF}\right) = s(ii)_{D\to SF} = \sum_{q=1}^{n} s(qi)_{D} \\ \left(S^{i\ \ \ell=D}_{\ j\neq i\ \ k=SF}\right) = s(ij)_{D\to SF} = 0 \end{cases} \quad \text{for } i,j,q\epsilon\{1,\cdots,n\} \qquad (27b)$$

Equations (24a,25a,25b), Equations (24b,26a,26b) and Equations (24c,27a,27b) describe, respectively, the bottom, middle and top three-dimensional matrixes within the four-dimensional structure in Figure 4, which corresponds to the impact multiplex $\left[S^{i\ \ell}_{\ j\ k}\right]^{multiplex}$ of size $\mathrm{n} \times \mathrm{n} \times 3 \times 3$. The next step is to identify the multiplex strongly connected component. We apply Tarjan's algorithm to the unfolded $\left[S^{i}_{\ j}\right]^{multiplex}_{unfolded}$ of size $3n \times 3n$ :

$$[S^i{}_j]^{multiplex}_{unfolded} = \begin{bmatrix} [s(ij)_{FI\to FI}] & [s(ij)_{FI\to SF}] & [s(ij)_{FI\to D}] \\ [s(ij)_{SF\to FI}] & [s(ij)_{SF\to SF}] & [s(ij)_{SF\to D}] \\ [s(ij)_{D\to FI}] & [s(ij)_{D\to SF}] & [s(ij)_{D\to D}] \end{bmatrix} \tag{28}$$

and identify the $m_{multiplex}$ number of banks that have nonzero structural impact on the multiplex. The parameter $p^{multiplex}_{\substack{min \\ connected}}$ in simulating the contagion process is evaluated over the multiplex strongly connected component $[S^i{}_j]^{multiplex}_{\substack{connected \\ unfolded}}$ of size $3m_{multiplex} \times 3m_{multiplex}$, and $C(j)^{multiplex}_{\substack{modified \\ connected}}$ are the modified own funds corresponding to this parameter. The unfolding $[S^i{}_j]^{multiplex}_{\substack{connected \\ unfolded}}$ is in the format from Equation (28), now for the $m_{multiplex}$ banks, and preserves the spectral properties of $[S^i{}_j{}^\ell{}_k]^{multiplex}_{connected}$ of size $m_{multiplex} \times m_{multiplex} \times 3 \times 3$. By analogy with the algorithm from Sections 3, now the Eigenpair $\left(\lambda^{max_unfolded}_{\substack{multiplex \\ connected}}, u^{unfolded}_{s\,\substack{multiplex \\ connected}}\right)$ of $[S^i{}_j]^{multiplex}_{\substack{connected \\ unfolded}}$ is generated. Then, the Eigenpair $\left(\lambda^{max}_{\substack{multiplex \\ connected}}, U_{s\,\substack{multiplex \\ connected}}\right)$ for $[S^i{}_j{}^\ell{}_k]^{multiplex}_{connected}$ is obtained as:

$$\lambda^{\max}_{\substack{multiplex \\ connected}} = \lambda^{\max_unfolded}_{\substack{multiplex \\ connected}} \quad \text{and} \quad U_{s\,\substack{multiplex \\ connected}} \overset{folding}{\Longleftarrow} u^{unfolded}_{s\,\substack{multiplex \\ connected}} \tag{29}$$

where $U_{s\,\substack{multiplex \\ connected}}$ is an Eigenmatrix of size $m_{multiplex} \times 3$ rather than an Eigenvector. Following the approach in Serguieva (2016, 2017a), we formulate the multiplex systemic risk $SR^{multiplex}_{risk}$ and resilience $SR^{multiplex}_{resilience}$ as:

$$SR^{multiplex}_{risk} = \begin{cases} \lambda^{max}_{\substack{multiplex \\ connected}} - p^{multiplex}_{\substack{min \\ connected}} > 0 \\ 0 \;,\; if \;\; \lambda^{max}_{\substack{multiplex \\ connected}} - p^{multiplex}_{\substack{min \\ connected}} \le 0 \end{cases} \tag{30a}$$

$$SR^{multiplex}_{resilience} = \begin{cases} 0 \;,\; if \;\; \lambda^{max}_{\substack{multiplex \\ connected}} - p^{multiplex}_{\substack{min \\ connected}} \ge 0 \\ p^{multiplex}_{\substack{min \\ connected}} - \lambda^{max}_{\substack{multiplex \\ connected}} < 0 \end{cases} \tag{30b}$$

The multiplex systemic-impact indexes are:

$$SII(i)^{multiplex}_{index} = \\ = \begin{cases} SII(i)^{multiplex}_{\substack{index \\ connected}} = \dfrac{r(i)_{s\,\substack{multiplex \\ connected}}}{\sum_{i=1}^{m_{multiplex}} \left( r(i)_{s\,\substack{multiplex \\ connected}} \right)} \;,\; \text{for } \; i \in \{1, \dots, m_{multiplex}\} \\ SII(i)^{multiplex}_{\substack{index \\ unconnected}} = 0 \;,\; \text{for } \; i \in \{m_{multiplex} + 1, \cdots n\} \end{cases} \tag{31a}$$

where:

$$r_{S_{connected}^{multiplex}} = U_{S_{connected}^{multiplex}} \; [1\;1\;1]' \tag{31b}$$

and the with corresponding surcharges are:

$$SIS(i)_{surcharge}^{multiplex} = \gamma_{EAD} \, SII(i)_{index}^{multiplex} =$$

$$= \begin{cases} SIS(i)_{\substack{surcharge \\ connected}}^{multiplex} = \gamma_{EAD} \, SII(i)_{\substack{index \\ connected}}^{multiplex} & \text{for } 0 < \gamma_{EAD} \ll 1; \; i \in \{1, \cdots, m_{multiplex}\} \\ SRS(i)_{\substack{surcharge \\ unconnected}}^{multiplex} = 0 & \text{for } \; i \in \{m_{multiplex} + 1, \cdots, n\} \end{cases} \tag{31c}$$

The multiplex stabilisation strategy is designed as follows. The parameter $\gamma_{multiplex}$ is optimised to estimate the minimum fractions of capital $\gamma_{multiplex} \; SII(i)_{\substack{index \\ connected}}^{multiplex} \; C(i)_{\substack{modified \\ connected}}^{multiplex}$ for each institution $i \in \{1, \dots, m_{multiplex}\}$ that when distributed in a balancing way among institutions $j \in \{1, \dots, m_{multiplex}\}$, in proportion to the impacts of $i$ within the multiplex, will bring the system to a targeted threshold $SR_{\substack{risk \\ threshold}}^{multiplex}(t)$ or $SR_{\substack{resilience \\ threshold}}^{multiplex}(t)$. The proportion is the ratio of the impact $\left(S{}^{i}{}_{j}{}^{\ell}{}_{k}\right)_{\substack{multiplex \\ connected}}$ of bank $i$ in market $\ell$ on bank $j$ in market $k$, divided by the overall impact $s(i)_{\substack{multiplex \\ connected}} = \sum_{z=1}^{3} \sum_{y=1}^{3} \sum_{q=1}^{m_{multiplex}} \left(S{}^{i}{}_{q}{}^{y}{}_{z}\right)_{\substack{multiplex \\ connected}}$ of bank $i$ within the multiple-market structure, for $i, j \in \{1, \cdots, m_{multiplex}\}$ and $\ell, k \in \{1,2,3\}$. Let the fraction that $j$ receives from $i$, in result of this, is denoted with $X(ij)$ . Then the non-charged four-dimensional matrix $S_{\substack{multiplex \\ connected}} = \left[\left(S{}^{i}{}_{j}{}^{\ell}{}_{k}\right)_{\substack{multiplex \\ connected}}\right]$ representing the multiplex is modified into the impact structure $S_{\substack{multiplex \\ connected}}^{rebalanced}$ as follows:

$$\left[\left(S{}^{i}{}_{j}{}^{\ell}{}_{k}\right)_{\substack{multiplex \\ connected}}^{rebalanced}\right] = \left[\left(S{}^{i}{}_{j}{}^{\ell}{}_{k}\right)_{\substack{multiplex \\ connected}} \Bigg/ \left(1 + \sum_{i=1}^{m_{multiplex}} \left(\frac{\sum_{\ell=1}^{3}\sum_{k=1}^{3}\left(X{}^{i}{}_{j}{}^{\ell}{}_{k}\right)}{p_{\substack{min \\ connected}}^{multiplex} \; C(j)_{\substack{modified \\ connected}}^{multiplex}}\right)\right)\right] =$$

$$= \left[\frac{\left(S{}^{i}{}_{j}{}^{\ell}{}_{k}\right)_{\substack{multiplex \\ connected}}}{1 + \sum_{i=1}^{m_{multiplex}} \left(\gamma_{multiplex} \; SII(i)_{\substack{index \\ connected}}^{multiplex} \left(\frac{C(i)_{\substack{modified \\ connected}}^{multiplex}}{p_{\substack{min \\ connected}}^{multiplex} \; C(j)_{\substack{modified \\ connected}}^{multiplex}}\right)\left(\frac{\sum_{\ell=1}^{3}\sum_{k=1}^{3}\left(S{}^{i}{}_{j}{}^{\ell}{}_{k}\right)_{\substack{multiplex \\ connected}}}{\sum_{z=1}^{3}\sum_{y=1}^{3}\sum_{q=1}^{m_{multiplex}}\left(S{}^{i}{}_{q}{}^{y}{}_{z}\right)_{\substack{multiplex \\ connected}}}\right)\right)}\right] \tag{32}$$

$$\text{for } 0 \le \gamma_{multiplex} \ll 1 \; ; \; i, j \in \{1, \cdots, m_{multiplex}\} \; ; \; \ell, k \in \{1,2,3\}$$

It considers that the funds $A(j)^{multiplex}_{connected} = C(j)^{multiplex}_{connected} - MC(j)^{multiplex}_{connected}$ available to $j$ increase to $A(j)^{rebalanced}_{\substack{multiplex\\connected}} = A(j)^{multiplex}_{connected} + \sum_{i=1}^{m_{multiplex}} X(ij)$ with the proportionate fractions $X(ij)$. We denote the ratio of available funds to total own funds of $j$ as $p(j)^{multiplex}_{connected} = a(j)^{multiplex}_{connected}\ p^{multiplex}_{\substack{min\\connected}} = \frac{A(j)^{multiplex}_{connected}}{C(j)^{multiplex}_{connected}} = \frac{a(j)^{multiplex}_{connected}\ A(j)^{multiplex}_{connected}}{C(j)^{multiplex}_{\substack{modified\\connected}}}$.

Maintaining the parameter $p^{multiplex}_{\substack{min\\connected}} = p^{\substack{multiplex\\rebalanced}}_{\substack{min\\connected}}$, for comparability of the simulation of contagion within the rebalanced structure leads to:

$$p^{multiplex}_{\substack{min\\connected}} = \frac{A(j)^{multiplex}_{connected}}{C(j)^{multiplex}_{\substack{modified\\connected}}} = \frac{A(j)^{\substack{multiplex\\rebalanced}}_{connected}}{C(j)^{\substack{multiplex\\rebalanced}}_{\substack{modified\\connected}}} = \frac{A(j)^{multiplex}_{connected} + \sum_{i=1}^{m_{multiplex}} X(ij)}{C(j)^{\substack{multiplex\\rebalanced}}_{\substack{modified\\connected}}} \tag{33a}$$

to a new modified value $C(j)^{\substack{multiplex\\rebalanced}}_{\substack{modified\\connected}}$ after rebalancing:

$$C(j)^{\substack{mutliplex\\rebalanced}}_{\substack{modified\\connected}} = C(j)^{multiplex}_{\substack{modified\\connected}} \left( 1 + \frac{\sum_{i=1}^{m_{multiplex}} X(ij)}{p^{multiplex}_{\substack{min\\connected}}\ C(j)^{multiplex}_{\substack{modified\\connected}}} \right) \tag{33b}$$

which produces the denominator in Equation (32).

The rebalancing reduces the largest Eigenvalue $\lambda^{\substack{max\\rebalanced}}_{\substack{multiplex\\connected}} < \lambda^{max}_{\substack{multiplex\\connected}}$, which is equivalent to building in structural resilience. Thus, the minimum redistribution pre-emptively reduces the effect of potential contagion in quarter $t$ based on the multiplex structure of exposures and the minimum capital requirements at the end of quarter $(t-1)$. The mechanism can be implemented automatically within the market infrastructure. It does not restrict the emerged preferences of banks for interaction within the multiplex of markets, but rebalances – at minimum cost and adaptively – how the system covers exposures collectively through the existing interlinkages. The mechanism also allows the banks to adapt their interaction preferences within the rebalancing impact structure, through incentives towards the emergence of a more resilient structure. In the terminology of computational intelligence approaches, this is analogous to the methodology of 'reinforcement learning'. The optimum mechanism involves not only the very top few but all reporting institutions that have nonzero systemic impact within the multiplex of markets at end of quarter $(t-1)$. The institutions are involved proportionately to their systemic impact at $(t-1)$, which is their potential to affect structural fragility and resilience in quarter $t$. The subtle rebalancing uses this potential and builds in resilience, instead of allowing this potential to drive the system further into fragility. The mechanism does not collect the surcharges into a fund to sit aside, but immediately uses them to achieve a stabilisation effect pre-emptively. Waiting for institutions to get in distress in order to access a fund will cost more. The redistribution also immediately compensates all institutions after the surcharges, where different institutions are compensated to a different extent. Thus effectively, each institutions is charged even less than the fraction of capital evaluated at the first step of the algorithm. While the charge depends on the systemic impact of a bank, its compensations depend on the systemic impact of other banks that affect the first bank through interlinkages. Finally, the potential for multichannel contagion through the multiplex structure contributes more to systemic fragility than single-channel contagion, however a positive point is that multichannel stabilisation also contributes more to systemic resilience than single-channel stabilisation.

## 4.2. Empirical evaluation

The empirical results presented in Tables 4 and 5 are evaluated for one of the quarters in the period from June 2014 to September 2015. Table 4 indicates that the multiplex structure does not meet the stability condition $\lambda^{max}_{\substack{multiplex \\ connected}} =< 0.1457$ , and therefore is in the region of structural fragility. The systemic risk of the unbalanced structure is $SR^{multiplex}_{risk} = 0.32867$ , and contagion will not be contained if triggered. If a threshold of $SR^{multiplex}_{\substack{risk \\ threshold}} = 0$ is targeted, then a stabilisation strategy with an optimum parameter $\gamma_{multiplex} = 0.02850$ will bring the system below this threshold.

### Structural Resilience of the Empirical Multiplex

based on data for one of the quarters in the period from June 2014 to September 2015

Table 4

| | |
|---|---|
| number of reporting banks | $n = 22$ |
| number of banks in the strongly connected subtensor | $m_{multiplex} = 19$ (18 overlapping banks with the derivative market) |
| $p^{multiplex}_{\substack{min \\ connected}}$ | 0.14573 |
| stability condition | $\lambda^{max}_{\substack{multiplex \\ connected}} < 0.14573$ |
| $SR^{multiplex}_{risk}$ ( no stabilisation implemented and $\lambda^{max}_{\substack{multiplex \\ connected}} = 0.47440$ at $\gamma_{multiplex} = 0$ ) | 0.32867 |
| $SR^{\substack{multiplex \\ rebalanced}}_{resilience}$ ( stabilisation implemented and $\lambda^{\substack{max \\ rebalanced}}_{\substack{multiplex \\ connected}} = 0.14568$ at $\gamma^{min}_{multiplex} = 0.02850$ ) | 0.00005 |

The structural resilience of the rebalanced system is $SR^{\substack{multiplex \\ rebalanced}}_{resilience} = 0.00005$, and contagion will be contained if triggered. The number of banks with nonzero systemic impact at the end of this quarter is 19, and they participate in the stabilisation step at the start of the next quarter. Notice that the threshold may be $SR^{multiplex}_{\substack{risk \\ threshold}} \neq 0$. Ffor example, within a long-term meta-strategy $0 < SR^{multiplex}_{\substack{risk \\ threshold}}(t) < SR^{multiplex}_{\substack{risk \\ threshold}}(t-1)$. Alternatively, $SR^{\substack{multiplex \\ rebalanced}}_{resilience} = 0.00005$ may be considered as too small. Though a potential contagion will be contained, a significant part of the system may be destroyed. Thus, a larger resilience threshold may be targeted $SR^{multiplex}_{\substack{resilience \\ threshold}} > 0.00005$.

Systemic Impact Ranking and Indexes
in Multiple Markets vs a Single Market

based on data for one of the quarters in the
period from June 2014 to September 2015

Table 5

| institutions | E | F | G |
|---|---|---|---|
| rank at $\gamma_{multiplex} = 0$, (multiple-market contagion dynamics) | 2 | 10 | 0 (not participating in the multiplex strongly-connected component) |
| rank at for $\gamma^{EAD}_{derivatives} = 0$, (single-market contagion dynamics) | 17 | 9 | 18 |
| $SII(i)^{multiplex}_{index}$ at $\gamma_{multiplex} = 0$ (multiple-market systemic impact) | 16.34% | 0.33% | 0% |
| $SII^{derivatives}_{EAD_index}(i)$ at $\gamma^{EAD}_{derivatives} = 0$ (single-market systemic impact) | 0.28% | 4.05% | 0.23% |

Table 5 presents the systemic impact indexes of three banks encoded as $E, F, G$. Institution $E$ has a high systemic impact in the multiplex and contributes significantly to multiple-market contagion and stabilisation. However, $E$ is of little importance in the single-layer structure of the derivatives market, and will contribute little to destabilising or stabilising processes there. Institution $F$ is of medium importance in both structures, but contributes different proportions to systemic risk (resilience) in the multiple-market system and in the single market. Bank $G$ has no systemic significance in multiplex contagion, while still contributing systemic impact in the single market. The empirical results show that banks differ in their significance and ability to influence the structure under the multiple-market scenario and the single-market scenario. The institutions will participate to a different extent in strategies to embed structural resilience under the two scenario. Stabilising the single market will not stabilise the multiplex of markets. Stabilising the multiplex will stabilise the single markets in the context of their interlinkages within the overall system.

We can improve the stabilisation analysis further and include that available funds $A(j)^{multiplex}_{connected}$ of bank $j$ decrease with $Q(j) = \gamma_{Q_multiplex}\, SII(j)^{multiplex}_{\substack{index\\connected}}\, C(j)^{multiplex}_{\substack{modified\\connected}}$ in proportion to the surcharge $SIS(j)^{multiplex}_{\substack{surcarge\\connected}}$ on $j$, along with increasing with the compensations $\sum_{i=1}^{m_{multiplex}} X(ij)$ in proportion to the surcharges $SIS(i)^{multiplex}_{\substack{surcarge\\connected}}$ on banks $i \in \{1, \cdots, m_{multiplex}\}$. Therefore, the contagion parameter is:

$$p^{multiplex}_{\substack{min\\connected}} = \frac{A(j)^{multiplex}_{connected}}{C(j)^{multiplex}_{\substack{modified\\connected}}} = \frac{A(j)^{\substack{multiplex\\rebalanced}}_{connected}}{C(j)^{\substack{multiplex\\rebalanced}}_{\substack{modified\\connected}}} = \frac{A(j)^{multiplex}_{connected} + \sum_{i=1}^{m_{multiplex}} X(ij) - Q(j)}{C(j)^{\substack{multiplex\\rebalanced}}_{\substack{modified\\connected}}} \tag{34a}$$

and the modified value for $C(j)^{\substack{multiplex\\rebalanced}}_{\substack{modified\\connected}}$ after rebalancing is:

$$C(j)^{\substack{multiplex\\rebalanced}}_{\substack{modified\\connected}} = C(j)^{multiplex}_{\substack{modified\\connected}} \left( 1 + \frac{\sum_{i=1}^{m_{multiplex}} X(ij) - Q(j)}{p^{multiplex}_{\substack{min\\connected}}\, C(j)^{multiplex}_{\substack{modified\\connected}}} \right) \tag{34b}$$

This transforms Equation (32) into:

$$\left[ \left( s^{i\ \ell}_{j\ k} \right)^{rebalanced}_{\substack{multiplex\\connected}} \right] = \left[ \left( s^{i\ \ell}_{j\ k} \right)_{\substack{multiplex\\connected}} \Big/ \left( 1 + \frac{\sum_{i=1}^{m_{multiplex}} X(ij) - Q(j)}{p^{multiplex}_{\substack{min\\connected}}\, C(j)^{multiplex}_{\substack{modified\\connected}}} \right) \right] =$$

$$\left[ \frac{\left( s^{i\ \ell}_{j\ k} \right)_{\substack{multiplex\\connected}}}{1 + \frac{\gamma_{Q_multiplex}}{p^{multiplex}_{\substack{min\\connected}}} \left( \sum_{i=1}^{m_{multiplex}} \left( SII(i)^{multiplex}_{\substack{index\\connected}} \frac{C(i)^{multiplex}_{\substack{modified\\connected}}}{C(j)^{multiplex}_{\substack{modified\\connected}}} \left( \frac{\sum_{\ell=1}^{3} \sum_{k=1}^{3} \left( s^{i\ \ell}_{j\ k} \right)_{\substack{multiplex\\connected}}}{\sum_{y,z=1}^{3} \sum_{q=1}^{mmultiplex} \left( s^{i\ y}_{q\ z} \right)_{\substack{multiplex\\connected}}} \right) \right) - SII(j)^{multiplex}_{\substack{index\\connected}} \right)} \right] \tag{35}$$

Due to limits on the period that access to data has been granted for this research, the empirical analysis here does not include simulating the strategy from Equation (35). We have instead simulated with synthetic data resembling characteristics of the empirical data, and observed how the process from Equation (35) changes the systemic risk of structure like $\left[ \left( s^{i\ \ell}_{j\ k} \right)^{rebalanced}_{\substack{multiplex\\connected}} \right]$ when the parameter $\gamma_{Q_multiplex}$ is at value $\gamma^{min}_{multiplex}$ optimising Equation (32). The results show that the built-in resilience $\Delta SR^{Q_multiplex}_{risk} = SR^{Q_multiplex}_{risk} \left( \gamma^{min}_{multiplex} \right) - SR^{\substack{Q_multiplex\\rebalanced}}_{risk} \left( \gamma^{min}_{multiplex} \right) > 0$ is at least 55% of $\Delta SR^{multiplex}_{risk} \left( \gamma^{min}_{multiplex} \right)$. Therefore, the larger part of the built-in resilience is preserved. Due to limits on the access to data, contagion and stabilisation processes have not been yet simulated for the fixed-income single market and the securities-financing single market, either. A detailed comparison is provided in Serguieva (2016, 2017a) of centralities across different quarters in 2014 and 2015, and across the three single markets, the non-interconnected and the interconnected multiplexes. The analysis there though addresses the exposure structure rather than the impact structure and contagion is not simulated. The Katrz-Bonacich centrality of the exposure structure differs for the interconnected multiplex, the non-interconnected multiplex, and for each single market. Our next task will be to analyse empirically the impact structure across markets and reporting quarters, and to account for both surcharges and compensations in the rebalanced structure. We anticipate a nonlinear effect in the results for the interconnected and non-interconnected multiplexes.

## V: Conclusions

Single-layer networks have now been adopted in modelling financial systems, however this task rather requires multilayer models, or interconnected multiplex networks as first approximation. There are few studies using non-interconnected multiplexes for modelling the structure of financial systems, and this has limitations in representing and analysing the complex system. The existing analyses also use the networks to represent but not affect the structure, and the approaches quite loosely follow regulatory requirements. We have identified gaps not addressed in current research, and then formulated solutions and provided empirical analysis.

There are powerful implementations of ensemble networks to non-financial domains. We touched on their ability to approach problems where single networks cannot cope, when evolving an ensemble and implementing to equity analysis in (Serguieva, Kalganova, 2002). The nature of the problem in focus here requires multilayer rather than ensemble networks, however we still address the capabilities of evolving networks as highly effective computational-intelligence techniques. Evolving an interconnected multiplex network through multiple periods allows not only modelling the multiple-market structure but also simulating strategies and suggesting meta-strategies for subtly affecting the structure towards building in targeted resilience. The hybrid approach can work with dynamic meta-strategies.[8]

The contributions in this study are as follows:

(i) The structure accounts for minimum capital requirements based on risk weighted assets.

(ii) The contagion model is formulated with an overall 'infection' (spreading) rate that allows for a unique spreading rate of each institution, both in single-market contagion and in multiple-market contagion.

(iii) The structure of the derivatives market accounts for positive net exposures in two directions between the same two institutions, due to different netting sets and enforceable netting agreements.

(iv) The derivatives market is analysed acknowledging that exposures on a going-concern basis (to a non-failed bank) and exposures at-default (to a failing bank) differ. The values of MtM net derivatives exposures after collateral and MtM net derivatives exposures at default are used, correspondingly.

[8] The dynamic meta-strategies provide incentives for the participants to adapt to and discover more-resilient structures but do not impose a particular structure. In computational intelligence terminology, this is a reinforcement learning technique.

(v) Systemic risk measures and systemic resilience measures are formulated, both for a single market and for the interconnected multiplex of markets. These are structural rather than monetary measures. However, the focus here is on building in structural resilience that then allows a system to sustain its associated monetary value.

(vi) Systemic impact indexes are formulated for each institution, both in a single market and within the multiple-market structure. The terminology 'systemic impact index' rather than 'systemic risk index' is used to indicate that the potential of an institution to affect the structure, though contributing to contagion processes, can also be used in strategies to contribute to stabilisation processes.

(vii) An interconnected multiplex network is formulated to model multichannel contagion within multiple markets. The model is based on a recent study in (Serguieva, 2016, 2017a) using the tensorial framework, where tensors of different rank are derived step by step with detailed interpretation within the systemic risk domain. Here, the derived model is used directly and implemented to analyse the structure that incorporates simultaneously but distinctly three interconnected markets – the fixed-income, securities-financing and derivatives markets.

(viii) Single-channel and multiple-channel stabilisation strategies are formulated that subtly and adaptively evolve the structure towards targeted thresholds of lower systemic risk or higher systemic resilience. The stabilisation mechanism works at a minimum cost for each institution and no cost for the system as a whole. It introduces subtle structural changes that do not restrict emerged interactions and preferences among institutions but rather balance how the system as a whole copes with the emerged structure of exposures. The mechanism could be implemented as part of the market infrastructure. This may also lead to institutions gradually adapting their preferences to the mechanism, and thus leading to the emergence of interactions underlying a more stable structure that would involve fewer and infrequent stabilisation steps.

(ix) All institutions that participate at the end of a period in the strongly connected component of the multilayer network, also have nonzero systemic impact indexes and the potential to affect the structure at the beginning of the next period. Only if the system does not meet a targeted threshold at the end of a period, a stabilisation step is applied at the beginning of the next period. It involves all institutions with nonzero systemic index rather than the very top few, in order to achieve effective rebalancing, where minimum charged fractions are immediately redistributed as compensations. If we look for an analogy, this mechanism may resemble the varying margin within the current clearance mechanisms. This also acknowledges that systemic risk is not entirely a fault of an institution but of the emerged structure.

(x) Empirical simulations of single-channel and multiple-channel contagion and stabilisation processes are performed using large granular databases now available to the Bank of England. The simulations confirm the ability of the multiplex network to capture contagion dynamics throughout multiple interconnected markets. The simulations also confirm the ability of the designed multilayer stabilisation strategies to pre-emptively build in structural resilience and reduce a potential contagion effect. The empirical systemic impact indexes for the same institutions differ within a single market and multiple markets, and therefore a strategy that builds in resilience within a single market will not stabilise the interconnected multiplex of markets. Building in resilience within the multiplex will stabilise the single markets in the context of their interlinkages within the overall structure.

Next, we will extend the current analysis comparatively across different quarterly periods, involving in each period the three markets first separately and then as an interconnected multiplex. We will further design, simulate and compare different multi-period meta-strategies with dynamic thresholds. Finally, the multichannel processes can be instantiated with more granular and higher frequency data. We anticipate confirming within the more dynamic setting, the current result that the potential for multichannel contagion through the multiplex structure contributes more to systemic fragility than single-channel contagion, but multichannel stabilisation also contributes more to systemic resilience than single-channel stabilisation.

## References

Battiston, S., Puliga, M., Kaushik, R., Tasca, P., Caldarelli, G. (2012) Debtrank: Too central to fail? Financial networks, the FED and systemic risk. *Scientific Reports*, 2(541): 1-6.

Bargigli, L., di Iasio, G., Infante, L., Lillo, F., Pierobone, F. (2015) The multiplex structure of interbank networks, *Quantitative Finance*, 15(4): 673-691.

Bholat, D. (2016) Modelling metadata in central banks, *European Central Bank Statistics Paper Series*, v 13.

Bholat, D. (2015) Big data and central banks. *Bank of England Quarterly Bulletin*, 55(1): 86-93.

Bholat, D. (2013) The future of central bank data. *Journal of Banking Regulation*, 14(3): 185-194.

Boccalettia, S., Bianconi, G., Criado, R., del Genio, C.I., Gómez-Gardeñes, J., Romance, M., Sendiña-Nadal, I., Wang, Z., Zanin, M. (2014) The structure and dynamics of multilayer networks, *Physics Reports*, 544(1): 1–122.

Chatelin, F. (2013) *Eigenvalues of Matrices*, 2nd Ed., Society for Industrial and Applied mathematics SIAM.

Cont, R., Moussa, A., Santos, E. (2013) Network structure and systemic risk in banking systems. In: Fouque, J.-P., Langsam, J. (Eds.) *Handbook on Systemic Risk*, Cambridge University Press, pp 327-367.

De Domenico, M., Solé-Ribalta, A., Omodei, E., Gómez, S., Arenas, A. (2015) Ranking in interconnected multilayer networks reveals versatile nodes. *Nature Communications*, 6(6868): 1-6.

De Domenico, M., Solé-Ribalta, A., Cozzo E., Kivelä, M., Moreno, Y., Porter, M., Gómez, S., Arenas, A. (2013) Mathematical formulation of multi-layer networks. *Physical Review X*, 3(041022): 1-15.

Furfine, C. (2003) Interbank Exposures: Quantifying the Risk of Contagion. *Journal of Money, Credit and Banking*. 35(1): 111-128.

Kivelä, M., Arenas, A., Barthelemy, M., Gleeson, J., Moreno, Y., Porter, M. (2014) Multilayer networks, *Journal of Complex Networks*, pp 1-69.

Langfield, S., Liu, Z., Ota, T. (2014) Mapping the UK interbank system. *Journal of Banking and Finance*, 45: 288–303.

Markose, S. (2012) Systemic risk from global financial derivatives: a network analysis of contagion and its mitigation with super-spreader tax. *IMF Working Paper Series*, WP/12/282.

Nise, N. (2011) *Control Systems Engineering*, 6th Ed., Wiley.

Poledna, S., Molina- Borboa, J.L., Martínez-Jaramillo, S., van der Leij, M., Thurner S. (2015) The multi-layer network nature of systemic risk and its implications for the costs of financial crises, Journal of Financial Stability, 20: 70-81.

Serguieva, A. (2017b) Multichannel contagion vs stabilisation in multiple interconnected financial markets [Presentation:1-22], *LSE Systemic Risk Centre Seminar,* London School of Economics, 13th February.

Serguieva, A. (2017a) A systematic approach to systemic risk, forthcoming in Bank of England Staff Working Papers. (and with Bholat, D., [Presentation:1-25] *8th Conf. of the Irving Fisher Committee on Central Bank Statistics: Implications of the New Financial Landscape*, Bank for International Settlements, 8th September 2016.)

Serguieva, A. (2016) *A Systematic Approach to Systemic Risk Analysis*, University College London mimeo, pp 1-100.

Serguieva, A. (2015) Big Data: opportunities and issues for central banks [Presentation:1-50], session on Central Bank Statistics: from data delivery to analytical value add, *Central Banking Spring Training Series*, training course handbook, 21st April.

Serguieva, A. (2013b) Systemic risk identification, modelling, analysis, and monitoring: an integrated approach. *Computational Engineering, Finance, and Science ArXiv*, arXiv1310.6486: 1-19. (and *Proceedings of the 2014 IEEE Conference on Computational Intelligence for Financial Engineering and Economics*, p viii; and seminar at the Office for Financial Research, US Treasury, September 2014)

Serguieva, A. (2013a) An evolving framework for systemic risk analysis. *Bank of England / University College London internal report,* 22nd February.

Serguieva, A. (2012) Computational intelligence and systemic risk models. [Presentation:1-64] *Workshop on Systemic Risk Assessment: Identification and Monitoring*. Bank of England Centre for Central Banking Studies, workshop handbook, 7th November.

Serguieva, A., Kalganova, T. (2002) A neuro-fuzzy-evolutionary classifier of low-risk investments, *Proceedings of the 2002 IEEE International Conference on Fuzzy Systems*, pp 997-1002, IEEE.